\documentclass{article}
\usepackage[a4paper, total={14cm, 23cm}]{geometry}

\usepackage[utf8]{inputenc}

\usepackage{soul} 
\usepackage{comment} 

\renewcommand{\baselinestretch}{1.0} 

\usepackage[bookmarksnumbered = true, colorlinks = true, linktocpage = true, bookmarksopen = true,
    citecolor=blue,%
    filecolor=blue,%
    linkcolor=blue,%
    urlcolor=blue,
	backref=page]{hyperref}
\usepackage{natbib} 

\usepackage{graphicx}
\usepackage{float}

\usepackage{amsfonts} 
\usepackage{amsmath}
\usepackage{bm}
\usepackage{mathtools}

\usepackage{tabu}

\usepackage[ruled,vlined]{algorithm2e}

\usepackage{epstopdf}
\AppendGraphicsExtensions{.tif}

\usepackage{color, colortbl}

\usepackage{ragged2e}

\begin{document}

\large 
    \begin{center}

    \vspace*{1cm}
    \LARGE
    \textbf{Beyond data: leveraging non-empirical information and expert knowledge in Bayesian model calibration}
    
    \vspace{0.5cm}
    \large{Sarah A. Vollert$^{1,2,*}$, Christopher Drovandi$^{1,2,3}$, Cailan Jeynes-Smith$^{2,4}$, Luz V. Pascal$^{1,2,5}$, \& Matthew P. Adams$^{1,2,6}$} \\

    \vspace{0.5cm}
    \normalsize
    $^1$Centre for Data Science, Queensland University of Technology, Brisbane, Australia. \\

    $^2$School of Mathematical Sciences, Queensland University of Technology, Brisbane, Australia.\\

    $^3$Centre of Excellence for the Mathematical Analysis of Cellular Systems, Queensland University of Technology, Brisbane, Australia

    $^4$Department of Paediatrics, The University of Tennessee Health Science Centre, Memphis, United States of America.\\

    $^5$Commonwealth Scientific and Industrial Research Organisation, Dutton Park, Brisbane, Australia. \\

    $^6$School of Chemical Engineering, The University of Queensland, St Lucia, Australia.\\

    $^*$Corresponding author. E-mail: sarah.vollert@hdr.qut.edu.au.
    
    \vspace{1.0cm}
  
   \end{center}

{\normalsize 
 
\textbf{Abstract: }Mathematical models connect theory with the real world through data, enabling us to interpret, understand, and predict complex phenomena. However, scientific knowledge often extends beyond what can be empirically measured, offering valuable insights into complex and uncertain systems. Here, we introduce a statistical framework for calibrating mathematical models using non-empirical information. Through examples in ecology, biology, and medicine, we demonstrate how expert knowledge, scientific theory, and qualitative observations can meaningfully constrain models. In each case, these non-empirical insights guide models toward more realistic dynamics and more informed predictions than empirical data alone could achieve. Now, our understanding of the systems represented by mathematical models is not limited by the data that can be obtained; they instead sit at the edge of scientific understanding.

\textbf{Teaser: }Expert knowledge, scientific theory, and observed phenomena can improve predictions and inference via model calibration.}

\newpage
\normalsize

\section{Introduction}
Modelling and simulation are invaluable tools for understanding complex systems across a wide range of disciplines, including biology and medicine \citep{villaverde_2014,Kitano_2002,qiao_2024}, ecology and environmental sciences \citep{Mouquet_2015,Geary_2020}, climate and weather science \citep{sillmann_2017}, social sciences \citep{Bellomo_2013,Holme_2015}, physical sciences \citep{sundberg_2012_astrophysics,wellmann_2018_geophysics}, and engineering \citep{sinha_2001_engineering}. Mathematical models -- such as differential equation models, or agent-based models -- encode system knowledge into mathematical frameworks, enabling scientists to analyse, understand and manage complex systems \citep{craver_2006}. However, these models become useful only if the values of uncertain parameters lead to realistic model simulations.

Model calibration (also referred to as parameterisation, parameter estimation, inverse problems or inference) is the process of fine-tuning model parameters to maximise the chance that model predictions are aligned with observations of the underlying system \citep{aster_2018_calibration}. During model calibration, we adjust parameter values to minimise discrepancies between model simulations and observations, thereby improving predictions \citep{Sisson_2018_ABC,mcelreath_2018,martin_2020}. Since model outputs are often highly sensitive to parameter choices \citep{saltelli_2006}, careful calibration is essential for ensuring accurate representations of the system. Additionally, inferred parameter values can enhance our understanding of underlying system mechanisms \citep{mcelreath_2018}, making calibration valuable not only for forecasting but also for scientific discovery. Thus, identifying parameter values that match empirical observations is a critical process for model development.

However, for many systems we study, obtaining data for calibration is challenging. While advances in technology and data collection have led to an explosion of available data, many scientific domains still face practical challenges that limit data availability. Data collection can be limited by accessibility (e.g., in deep-sea \citep{levin_2019_deepsea}, or active volcanoes \citep{deligne_2018_volcano}), large scales (e.g., ecosystem species monitoring; \cite{Geary_2020}), or invasiveness (e.g., brain biopsies for disease research; \cite{malone_2015_brainbiopsy}). Time constraints further complicate data availability, especially if events are rare (e.g., floods; \cite{assumpccao_2018_flood}), gradual (e.g., climate change; \cite{karl_1989_climate}) or emerging (e.g., pandemics; \cite{britton_2019_emergingvirus}). Moreover, some model outputs are inherently unmeasurable and rely on uncertain proxies, such as subjective quality-of-life indicators \citep{felce_1995_quality_of_life}. While big data has revolutionised fields where continuous or automated data collection is possible, the data in many critical modelling problems remains low quality, low quantity, or high cost. In these cases, modellers must use what data (if any) they can get. Limited data availability diminishes the precision of model calibration, as there may be a wide range of plausible parameter values consistent with the data, and this high uncertainty in parameter values can propagate through to high uncertainty in model predictions. This is especially problematic if the value of a parameter controlling a positive or negative effect on model behaviour is left highly uninformed (see e.g., \cite{botelho_2024}). The effects of limited data on inferences and predictions can reduce trust in models and their ability to inform, analyse and explore a system \citep{harper_2021}. 

Yet, even in the absence of classically observed data, many systems that we model have been extensively studied. This research can yield a wealth of information that may not conform to what we classically think of as data -- information such as system understanding, documented phenomena, observational insights, and expert knowledge. As such, here we distinguish two types of information. \textit{Empirically measured information} refers to data obtained through direct measurement or observation using systematic, reproducible, and objective methods, such as data collected from experiments, field studies, and sensor recordings. In contrast, we define \textit{non-empirical information} as any knowledge that can inform or constrain a model's outputs but is not directly measured and, therefore, is not classified as empirically measured information (see Table \ref{tab:Non-empirical examples} for examples). Beyond specific examples of non-empirical information, such as those in Table \ref{tab:Non-empirical examples}, expert-elicited knowledge can be obtained for calibration in any well-studied system \citep{krueger_2012}. Experts tend to intimately understand the systems they research. For example, if an expert is shown model predictions, and they can identify areas where the predictions appear ``wrong" then this information can be utilised in model calibration as non-empirical information. 

In this paper, we develop a comprehensive statistical framework that can simultaneously integrate all kinds of empirical and non-empirical information for model calibration in a statistically rigorous manner. This approach opens up new avenues for model calibration by harnessing the power of often-overlooked non-empirical information -- information that is usually not considered as data. To accomplish the calibration of simulation models to both empirical and non-empirical information, we formalise this process in an approximate Bayesian framework. Approximate Bayesian computation (ABC) is a flexible approach for model calibration that matches summaries of the model simulations to that of an empirical dataset \citep{Sisson_2018_ABC,Beaumont_2019_ABC,sunnaaker_2013_ABC}. In the case of non-empirical information, we show that summary statistics of an empirical dataset can be substituted for elicited summary statistics, informed by expert knowledge, observed phenomena, theories or other non-empirical information. While ABC was designed for calibrating models to empirical datasets \citep{Sisson_2018_ABC,sunnaaker_2013_ABC}, and few studies have considered non-empirical constraints \citep{barnes_2011,vollert_2024_alternativeConstraints}, to the authors' best knowledge, the present work is the first formal quantitative framework for calibrating models to both. Efficiently obtaining model parameter sets from a combination of any empirical and non-empirical information presents an interesting computational challenge. In response, we used sequential Monte Carlo (SMC) techniques \citep{delmoral_2006,chopin_2002,drovandi_2011_ABC} to develop a bespoke sampling algorithm that simultaneously combines the empirical data and summary statistics of the non-empirical information in an efficient, robust and flexible manner.

\begin{table}[ht!]
    \centering
    \begin{tabular}{p{0.19\linewidth} p{0.27\linewidth} p{0.23\linewidth} p{0.16\linewidth}}
        \hline \hline \multicolumn{4}{c}{\textbf{Non-empirical information}} \\ \hline
        \textbf{Scientific \qquad discipline} & \textbf{Original statement} & \textbf{Effect on model calibration} & \textbf{References} \\ \hline \hline
        Biogeochemistry & Lakes can be either oligotrophic or eutrophic and remain in that state until substantially perturbed. & There should be two alternative stable steady states of the nutrient concentration. & {\footnotesize\cite{bhagowati_2019,carpenter_1999_lake}}\\ \hline
        Biology & Multiple sclerosis exhibits relapsing-remitting symptoms that correspond with lesion development. & Modelled degeneration should oscillate as it progresses. & {\footnotesize\cite{confavreux_2000,mendizabal_2011_MS}} \\ \hline
        Climate \qquad \qquad science & Ice core samples can indicate climatic trends across long periods. & Global energy balance models must reflect observed long-term climatic trends. & {\footnotesize\cite{cuffey_2000_ice,dommenget_2011_EBM}} \\ \hline 
        Ecology & Introducing pigs to the California Channel Islands led to an increase in golden eagle populations and a decrease in fox populations. & When historical perturbations are simulated, predicted population responses must reflect the observed trends.  & {\footnotesize\cite{roemer_2002_unintendedConseq,vollert_2024_alternativeConstraints}} \\ \hline
        Environmental science & Locals who have experienced a flooding event know which roads were submerged and at what times. & Modelled water depth should align with highly localised information. & {\footnotesize\cite{saunders_2025,fava_2019_flood,assumpccao_2018_flood}} \\ \hline
        Epidemiology & Outbreaks of COVID-19 in various countries were typically followed by a second smaller wave of infections due to waning immunity. & There should be two consecutive peaks in infections, where the second is smaller. & {\footnotesize\cite{rachel_2024_covid,friston_2020_covid}} \\ \hline
        Sports \qquad \qquad science & Athletic performance is limited by what can be physiologically achieved, such as heart rate maxima. & Specific quantities of interest are restricted by physiologically reasonable bounds. & {\footnotesize\cite{lundby_2015_sport,stessens_2024_sport}} \\ \hline \hline
    \end{tabular}
    \caption{Examples of non-empirical information that can be used for calibrating models. }
    \label{tab:Non-empirical examples}
\end{table}

From a Bayesian perspective, our method provides a new way to define informed prior distributions using simulation. Prior elicitation is the process of formally translating expert knowledge into a prior distribution for parameter values, which is crucial for reliable calibration and inference \citep{mikkola_2024, Banner_2020, gelman_2017}. Our approach offers a generalisable and robust alternative prior elicitation method, enabling the construction of informed prior distributions with the added benefit of flexible parameter dependency structures -- a non-trivial task \citep{gelman_2017, mikkola_2024}.

While the framework we propose presents several statistical advances, we focus on the benefits of these methods for mechanistic modellers. Calibrating models with non-empirical information can ensure that a model used for decision-making or analysis contains the behaviours exhibited in the real system, even if the dataset does not. We present several case studies that illustrate the benefits of integrating non-empirical features into model calibration. Firstly, we calibrate a logistic growth model forecasting coral populations using estimates of recovery time post-disturbance (\nameref{logistic case study}). Then, we calibrate an ecological four-species dynamic population model using both a sparse time-series dataset and principles from theoretical ecology governing the expected equilibrium behaviour of the system (\nameref{ecology case study}), demonstrating that ignoring long-standing ecological theory can change the predicted outcomes of conservation and management decisions. Lastly, we calibrate a biochemical network model using the observation that these systems must both respond and adapt to changes in the stimulus (\nameref{cellular adaptation case study}), revealing new insights into the binding affinity of catalysts driving homeostasis, relevant for pathology. Together, these case studies showcase the profound effect of non-empirical information on model predictions, management implications, and system understanding. 

\section{Results}

\subsection{Case study 1: Logistic coral growth}
\label{logistic case study}
Firstly, we demonstrate our approach with a simple logistic growth example. The logistic growth model is ubiquitous in many areas of ecology, medicine, and biology \citep{Murray_2002,simpson_2022_sigmoidReefs}. Here, we will use the logistic growth model to represent the recovery dynamics of coral populations. Using logistic growth, the percentage of coral cover on a reef can be modelled as
\begin{equation}
    \frac{\rm{d}\mathit{y}}{\rm{d}\mathit{t}} = ry(t)\left(1-\frac{y(t)}{K}\right),\qquad y(0) = y_0,
\end{equation}
where $y$ is the coral cover (\% area), $t$ is the time (years), $r$ is a growth rate parameter (year$^{-1}$), $K$ is the carrying capacity (\% area), and $y_0$ is the initial coral cover (\% area). This ordinary differential equation can be solved analytically as 
\begin{equation}
    y(t) = \frac{K y_0}{y_0+(K-y_0)e^{-rt}}.
\end{equation}
Here, three model parameters may require estimation: the growth rate $r$, the maximum coral cover $K$, and the initial coral cover $y_0$. 

Other than time-series data, some information may be available to inform the values of parameters $r, K$ and $y_0$. For example, there may be physical restrictions on the carrying capacity $K$ of coral cover, physiological maxima for coral growth rates $r$, and knowledge of post-disturbance coral cover $y_0$. We use this information to define the prior distribution for model parameters, which describes the distribution of parameter probabilities before considering the data (e.g., $K \sim \mathcal{U}(60\%, 80\%)$; see Supplementary Materials Section \ref{SM: logistic details} for more details). However, even after capturing all prior knowledge of these parameters, they can be highly uncertain, and therefore of little value for ecological forecasting (see Figure \ref{fig:logistic_constraints_only}B, grey). In a typical parameterisation framework, time-series data would be used to infer these parameters' values (e.g.,\ \cite{simpson_2022_sigmoidReefs}), yielding a posterior distribution describing the probability of parameters after considering both the prior knowledge and the data. However, there may be no monitoring program on the reef of interest. For instance, the Australian Institute of Marine Science's Long Term Monitoring Program is one of the most comprehensive records of coral status on any reef ecosystem; yet with over $3000$ reefs on the Great Barrier Reef in Australia, less than 20\% have been surveyed \citep{AIMS_LTMP}. Hence, for the all-too-likely scenario that there is no data on a reef of interest, we instead demonstrate how the model could be calibrated using non-empirical information -- specifically, the elicited features of coral population recovery. 

\subsubsection{Expert elicited observations are easily incorporated into model calibration}
\label{Sec: non-empirical only logistic}

\begin{figure}[h]
    \centering
    \includegraphics[width=0.9\linewidth]{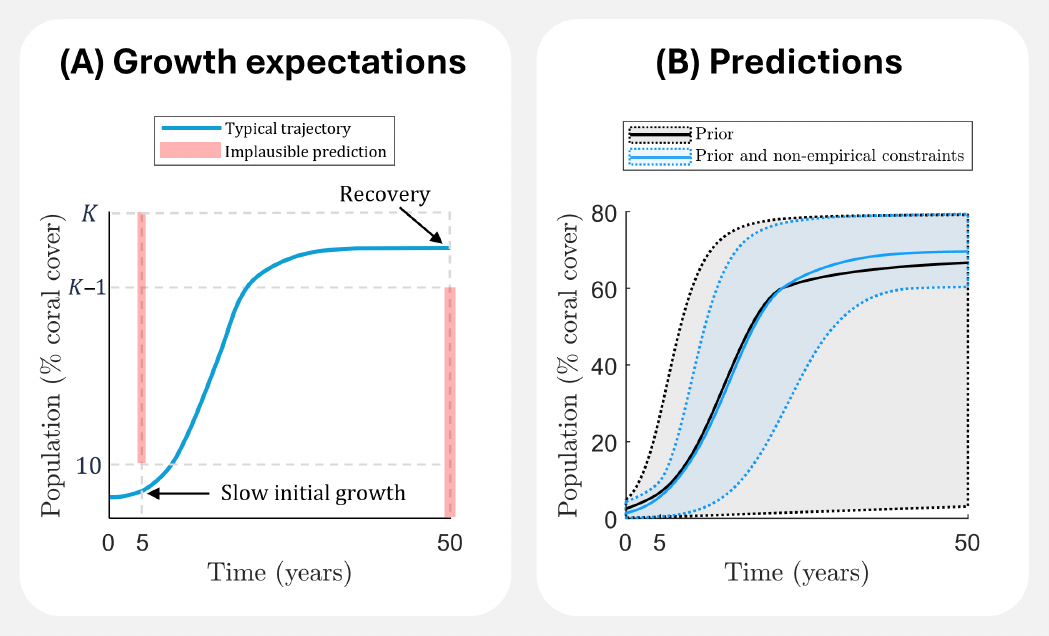}
    \caption{\textbf{(A)} Two non-empirical constraints on the growth of coral cover through time. Models predicting this growth should exhibit the two desired behaviours: slow initial growth (coral cover less than 10\% at 5 years) and recovery within 50 years (coral cover within 1\% area of carrying capacity $K$). \textbf{(B)} Model forecasts generated using ensembles of model parameters from both the prior distribution (grey) and the distribution constrained by both the prior and non-empirical sources (blue). Here, the median and 95\% credible intervals for predictions are shown as solid lines (black and blue) and shaded regions (grey and light blue) respectively. Models parameterised using both the prior and the non-empirical data sources exhibit the two desired behaviours: slow initial growth and recovery at 50 years. Notice that incorporating these non-empirical constraints significantly shifts the 95\% credible interval away from undesirable model behaviours.}
    \label{fig:logistic_constraints_only}
\end{figure}

For this illustrative example, we consider two non-empirical constraints. These constraints help ensure that model predictions do not exhibit behaviours deemed impossible by an expert. Following a disturbance (such as a marine heatwave), an ecologist may expect both slow initial growth of coral populations and full recovery within a specified time frame (see Figure \ref{fig:logistic_constraints_only}A). For example, for this reef it might be estimated that:
\begin{enumerate}
    \item ``Coral cover should be less than 10\% within the first 5 years''; and 
    \item ``Coral populations should be fully recovered (within 1\% area of carrying capacity) within 50 years.'' 
\end{enumerate}

These two statements can be expressed in terms of the model parameters $r, K$ and $y_0$:
\begin{align}
\label{eq:logistic_summary1}
    y(5) &= \frac{K y_0}{y_0+(K-y_0)e^{-5r}} \leq 10\%, \\
    \label{eq:logistic_summary2}
    y(50) &= \frac{K y_0}{y_0+(K-y_0)e^{-50r}} \geq K-1\%.
\end{align} 
However, it is still difficult to directly incorporate these two mathematical statements into the prior distribution. For instance, it would be challenging to describe how the prior distribution for $K \sim \mathcal{U}(60\%,80\%)$ should be updated to meet these conditions, both because there are multiple constraints to consider and because there are dependencies between the parameters. Instead, we treat $y(5)$ and $y(50)$ as summary statistics and construct non-empirical discrepancy functions $\rho_5$ and $\rho_{50}$ that represent the conditions in Equations \eqref{eq:logistic_summary1} and \eqref{eq:logistic_summary2}, which can be calculated for any given values of $r, K,$ and $y_0$. These summary statistics can then be easily compared to the behaviours that experts expect via one discrepancy function $\rho$: 
\begin{align}
    \rho &= \rho_{5} + \rho_{50},     \label{eq:logistic_discrepancy}\\
    \rho_{5} &= \mathrm{max}(0, y(5) - 10), \\
    \rho_{50} &= \mathrm{max}(0, K -1 - y(50)).    \label{eq:logistic_discrepancy_end}
\end{align}
In Equations \eqref{eq:logistic_discrepancy}-\eqref{eq:logistic_discrepancy_end}, $\rho_{5}$ measures the discrepancy of coral cover exceeding 10\% in the first 5 years, $\rho_{50}$ measures the discrepancy of coral cover not recovering within 1\% of $K$ in 50 years, and $\rho$ measures the total discrepancy between the simulated and expected coral cover.  

Using these informative non-empirical constraints together with the less informative prior distribution for parameters $y_0, K$ and $r$ (Figure \ref{fig:logistic_constraints_only}A), we generated an ensemble of parameter sets yielding model predictions that meet the expected behaviours. The incorporation of expert-elicited observations dramatically shifts the range of model predictions, from highly uncertain (Figure \ref{fig:logistic_constraints_only}B, grey), towards a more constrained range of plausible model behaviours (Figure \ref{fig:logistic_constraints_only}B, blue). The effect of each of the two individual non-empirical constraints on model predictions can also be clearly identified: the early model predictions are restricted by the slow initial growth constraint, and the predictions beyond five years are controlled by the long-term recovery constraint (Figure \ref{fig:logistic_nonempirical_prediction_comparison}). 

This ensemble generation can proceed via any appropriate ABC sampling algorithm \citep{Sisson_2018_ABC}, such as rejection sampling, if the non-empirical constraints are not particularly restrictive. However, throughout this work, we instead use an SMC-ABC algorithm (see Supplementary Material \ref{SM: sampler} for further details; \cite{drovandi_2011_ABC}) as it is an efficient sampling algorithm even when drawing proposals from the prior yields a low acceptance rate. SMC algorithms in general are also particularly robust to multi-modal parameter distributions \citep{delmoral_2012}. In the present work, we will refer to the combination of the prior distribution and \textit{any} constraints (whether non-empirical, a dataset, or both) as the posterior distribution.

\subsubsection{Incorporating constraints via simulation accounts for parameter interdependencies}

\begin{figure}[H]
    \centering
    \includegraphics[width=0.7\linewidth]{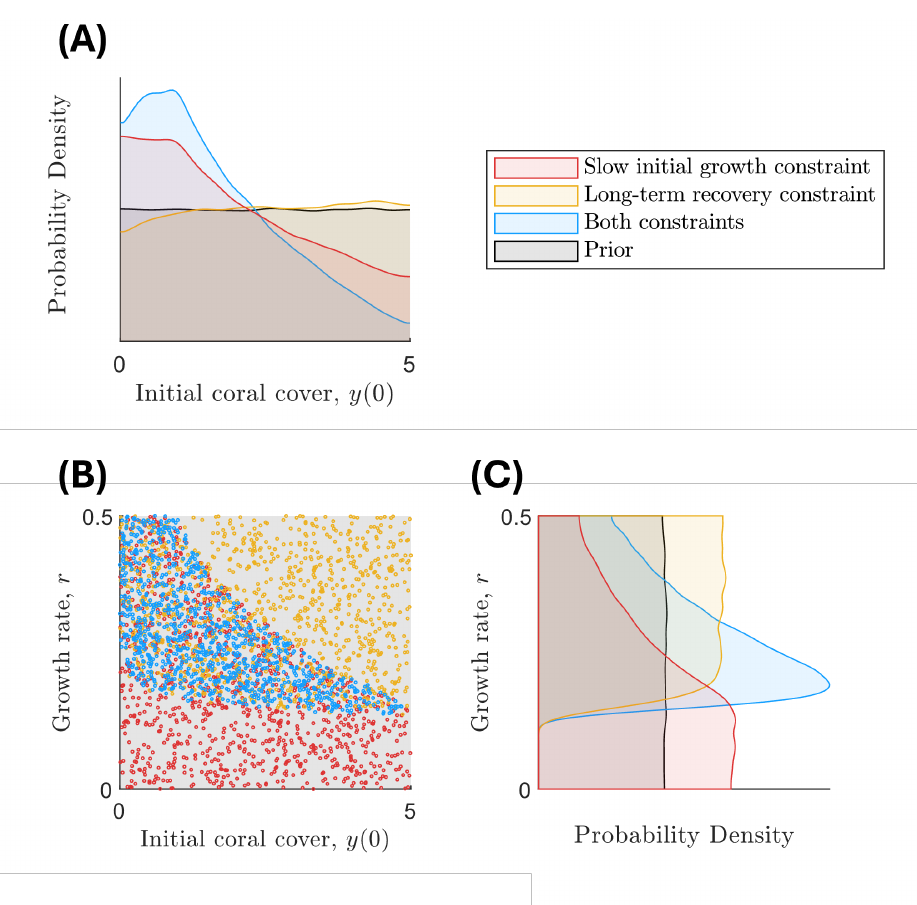}
    \caption{The marginal (A,C) and bivariate (B) parameter distributions for logistic growth model parameters $y_0$ and $r$ when sampled from the prior (grey), prior and non-empirical 5-year constraint (red), prior and non-empirical 50-year constraint (yellow), and prior and both non-empirical constraints (blue). This figure shows that the effects of constraints on individual parameters can be obfuscated if parameter interdependencies are not considered. Note that for the parameter $y_0$ (A), combining the two individual non-empirical constraints (red and yellow) does not yield a joint distribution (blue) between the individual distributions. This counter-intuitive effect can be understood by looking at the joint probability distribution (B) between parameters $y_0$ and $r$ obtained via simulation. }
    \label{fig:logistic_bivariate}
\end{figure}

From a prior elicitation perspective, we incorporate non-empirical information into the model via simulation because it is challenging to manually update the prior distribution to reflect these constraints. In this case, there are two key reasons that make it difficult to manually define a prior that matches the expert information: (1) the need to satisfy multiple constraints means there can be a trade-off in objectives that changing a parameter value must balance, and (2) mechanistic models create a parameter dependency structure such that parameters need to be jointly considered. Our results demonstrate how these challenges can arise even in this simple three-parameter model. 

In this logistic growth model, the value of the initial coral cover parameter $y_0$ needs to be small to ensure the model exhibits slow initial growth (Figure \ref{fig:logistic_bivariate}A, red shaded region; Equation \eqref{eq:logistic_summary1}), yet long-term recovery is less likely for small initial populations $y_0$ (Figure \ref{fig:logistic_bivariate}A, yellow shaded region; Equation \eqref{eq:logistic_summary2}). It might be expected that the combination of these constraints will balance the probability distribution of $y_0$ such that it sits between the two individual constraint distributions. Instead, the initial coral cover $y_0$ is even more likely to be a small value when both constraints are considered (Figure \ref{fig:logistic_bivariate}A, blue shaded region) due to compensatory effects with the growth rate parameter $r$. Since the growth rate $r$ must be higher to ensure long-term recovery (Figure \ref{fig:logistic_bivariate}C), there is less chance that the initial coral cover $y_0$ can be large and still meet the slow initial growth condition (Figure \ref{fig:logistic_bivariate}B). Hence, even for a simple three-parameter model, capturing two straightforward constraints (Equations \eqref{eq:logistic_summary1} and \eqref{eq:logistic_summary2}) yields non-trivial interdependencies between model parameters. Yet, generating the joint parameter distribution that captures these constraints via simulation effortlessly accounts for the model's parameter dependency structure, which is highly advantageous for prior elicitation. 

\subsubsection{Combining data with non-empirical observations can enhance predictions}
\begin{figure}[H]
    \centering
    \includegraphics[width=0.6\linewidth]{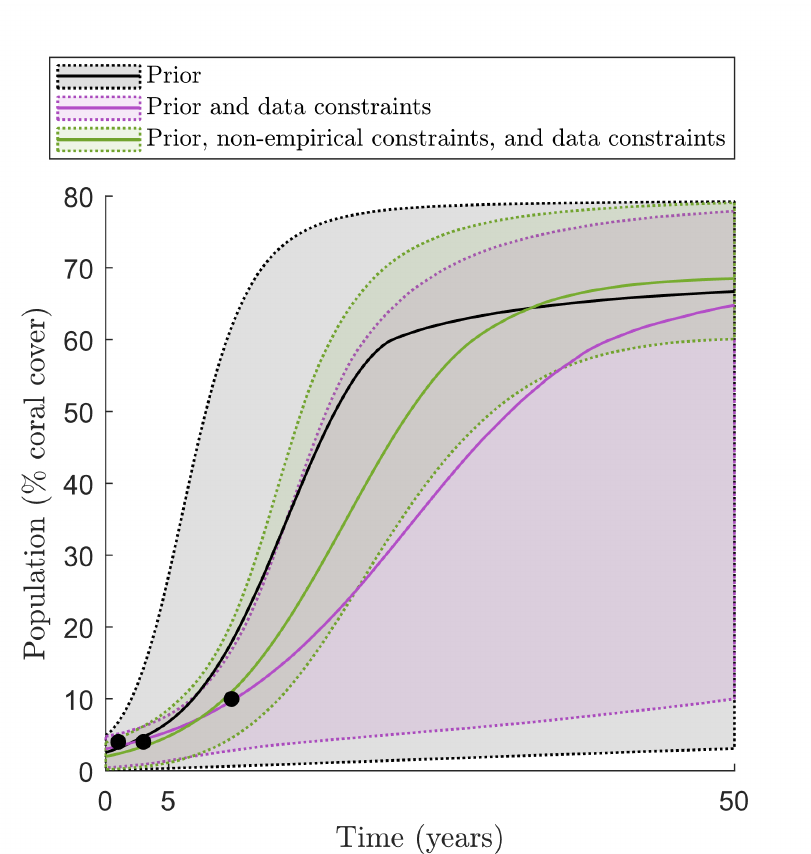}
    \caption{Predictions (median and 95\% credible intervals) generated from a prior distribution (grey), a posterior distribution constrained using the dataset (purple), and a posterior distribution constrained using data and non-empirical information, i.e. Equations \eqref{eq:logistic_summary1} and \eqref{eq:logistic_summary2} (green). Notice that both informed distributions (purple and green) match the dataset, but the one that is additionally constrained by non-empirical information (green) also meets the requirement for recovery within 50 years.}
    \label{fig:Logistic_predictions_data}
\end{figure}

While it may be common for a reef of interest to have no monitoring data, roughly $500$ out of more than $3000$ reefs on the Great Barrier Reef \textit{have} been surveyed \citep{AIMS_LTMP}, even though only $232$ reefs were surveyed more than three times \citep{AIMS_LTMP_dataset}. Hence, we next illustrate the effect of time-series data on model calibration for a hypothetical reef, using three sparse observations of coral cover. For these three data points, we use a Gaussian likelihood function to describe the probability that the model defined by a set of parameters produced the observed dataset (see Methods section \ref{new methods} or Supplementary Material section \ref{SM: logistic details} for more detail). Hence, this model-data fitting exercise accounts for the prior distribution, and the time-series data consisting of three observations simultaneously. We also show the outputs of model-data fitting that accounts for the prior distribution, the time-series data, and the non-empirical constraints simultaneously (Figure \ref{fig:Logistic_predictions_data}).

Combining limited data with non-empirical observations leads to model outputs that match both the data and the expert beliefs (Figure \ref{fig:Logistic_predictions_data}, green). Calibrating the model to the dataset (Figure \ref{fig:Logistic_predictions_data}, purple) can constrain the predictions to a more sensible domain in comparison to the prior distribution (Figure \ref{fig:Logistic_predictions_data}, grey). However, this posterior still allows model predictions that exhibit very little recovery in coral cover, which conflicts with the expert belief that coral cover must be within 1\% area of carrying capacity after 50 years. Instead, if the model is constrained also by the two non-empirical constraints that represent the expert beliefs (Section \ref{Sec: non-empirical only logistic}), the parameter inferences (Figure \ref{fig:Logistic_marginals_data}) and model outputs (Figure \ref{fig:Logistic_predictions_data}) satisfactorily match both the expert-informed requirements and the dataset.

\subsection{Case study 2: Ecosystem population modelling}
\label{ecology case study}

Ecosystem network models describe the interactions between species and simulate the dynamics of ecosystem populations over time. Ecosystem network models can provide broad insights into how all ecosystems function \citep{Allesina_2012_stab,Grilli_2017_FS,landi_2018} and forecast the potential effects of human impacts and conservation management on specific ecosystems of interest \citep{baker_2019,baker2017_EEM,adams2020_TS}. These ecosystem network models can be calibrated to match any available time-series data from ecological monitoring programs for a specific ecosystem \citep{adams2020_TS,baker_2019}. In addition, ecological theory suggests that the equilibrium species populations stably coexist in general \citep{Cuddington_2001, Allesina_2012_stab,Rohr_2014,Grilli_2017_FS,Dougoud_2018,landi_2018}; in other words, species populations can fluctuate over the short-term, but have stable long-term dynamics. Consequently, ecosystem network models for a specific ecosystem are often parameterised under the assumption that they exhibit a stable, coexisting equilibrium \citep{baker2017_EEM,Rendall_2021_EEMeg,Pesendorfer_2018_egEEM,Peterson_2021,Peterson_2021_DirkHartog,Bode_2017}. 

Approximate Bayesian methods have been previously used to parameterise theoretical beliefs about ecosystem equilibria \citep{baker2017_EEM,vollert_2023_SMCEEM,pascal_2024}, and a larger variety of methods have been used to calibrate these models to time-series data \citep{baker_2019,botelho_2024} including SMC algorithms \citep{adams2020_TS}. Previously, when equilibrium constraints and time-series data have both been fitted to ecosystem network models, the parameter space was searched using an ad-hoc method similar to SMC-ABC with iteratively introduced dynamic constraints underpinned by a distance measure \citep{baker_2019}. Here, we provide a robust and efficient method for calibrating ecosystem network models using equilibrium constraints and time-series datasets via our combined SMC algorithm (see Supplementary Material section \ref{SM: sampler} for more detail).

\begin{figure}[H]
    \centering
    \includegraphics[width=\linewidth]{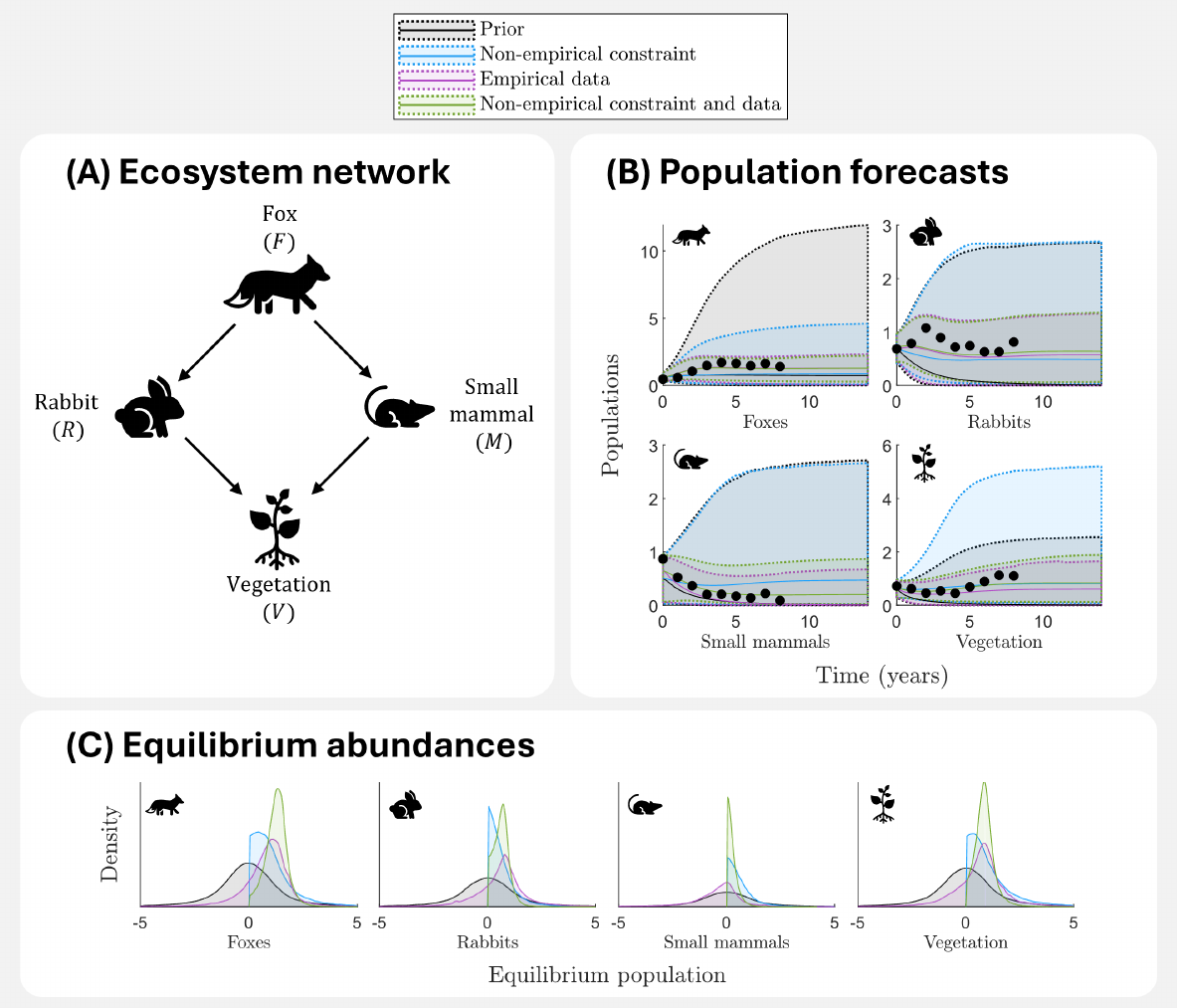}
    \caption{\textbf{(A)} Ecosystem network diagram where the arrows indicate the direction of the energy transfer associated with the species interaction due to predation or herbivory. \textbf{(B)} Predictions (median and 95\% credible intervals) generated from four distributions: the prior (grey), equilibrium-constrained posterior (blue), time-series data posterior (purple) and the combined equilibrium-constrained and time-series data posterior (green).  \textbf{(C)} Equilibrium populations for each species when calculated using parameter sets from each of the four distributions: the prior (grey), equilibrium-constrained posterior (blue), time-series data-informed posterior (purple), and the combined equilibrium-constrained and time-series data-informed posterior (green). Note, negative equilibrium abundances for a species do not imply that the specific species will go extinct, and instead suggest that not all species can coexist.}
    \label{fig:Eco_main_figure}
\end{figure}

In this case study, we consider a well-known four-species ecosystem network motif \citep{Monsalve_2022,Pech_1998_foxEcosystem} comprised of foxes ($F$), rabbits ($R$), small mammals ($M$) and vegetation ($V$), as depicted in Figure \ref{fig:Eco_main_figure}A. Using the ubiquitous Lotka-Volterra equations (see e.g., \cite{adams2020_TS,baker2017_EEM}), we converted this ecosystem network into a system of ordinary differential equations for calibration.

Firstly, we consider two non-empirical observations from ecological theory that describe the long-term behaviours of ecosystem populations: coexistence and stability. \textit{Coexistence} (often referred to as feasibility) specifically requires equilibrium populations for all species ($n_F^*, n_R^*, n_M^*,$ and $n_V^*$) to be positive \citep{Grilli_2017_FS}. \textit{Stability} is the ability of ecosystem populations to resist changes from small external pressures \citep{Allesina_2012_stab}, such that the real parts of all eigenvalues of the Jacobian matrix evaluated at equilibrium ($\mathbb{R}({\lambda_1}), \mathbb{R}({\lambda_2}), \mathbb{R}({\lambda_3}),$ and $\mathbb{R}({\lambda_4})$) must be negative (see \cite{vollert_2023_SMCEEM} for further details). For this four-species ecosystem network model, all eight summary statistics (four equilibrium abundances $n_F^*, n_R^*, n_M^*,$ and $n_V^*$, and four real components of eigenvalues $\mathbb{R}({\lambda_1}), \mathbb{R}({\lambda_2}), \mathbb{R}({\lambda_3}),$ and $\mathbb{R}({\lambda_4})$) are combined within a single discrepancy function $\rho$, calculated as 
\begin{align}
    \rho &=  \sum_{i=\{F,R,M,V\}} \big| \min \{ 0,n_i^*) \} \big|  + \sum_{i=j}^4 \big| \max\{0,\mathbb{R}(\lambda_j)\} \big|,
\end{align}
where $\rho$ measures the discrepancy in equilibrium behaviour from the theorised coexisting and stable behaviour \citep{vollert_2024_alternativeConstraints}.

In addition, we used simulated time-series data for model calibration, where each species has yearly abundance estimates. We used a Gaussian likelihood function which assumes normally distributed measurement noise to match the ecosystem network model's population prediction to the time-series data. Further detail on the model, prior distribution, summary statistics, and likelihood can be found in Supplementary Material section \ref{SM: ecosystem details}. 

\subsubsection{Non-empirical constraints can ensure long-term model behaviours are theoretically sound}

We generated an ensemble of 100,000 parameter sets for the non-empirical equilibrium constraints using SMC-ABC (see \cite{vollert_2023_SMCEEM}), for the simulated time-series data using SMC (see \cite{adams2020_TS}), and for both simultaneously using our combined SMC algorithm (see Supplementary Material \ref{SM: sampler}; marginal parameter distributions shown in Figure \ref{fig:ecol marginal}). Calibrating the model to the time-series data hones the predictions onto appropriate population sizes, and ensures that the trends in the data are captured (Figure \ref{fig:Eco_main_figure}B, purple). The long-term non-empirical constraints are required to ensure that the system's equilibrium matches the ecological theory (Figure \ref{fig:Eco_main_figure}C, blue; Figure \ref{fig:ecol stability distribution}). 

The combination of these two information sources (time-series data and non-empirical constraints) leads to models that both match the dataset well (Figure \ref{fig:Eco_main_figure}B, green) and exhibit the expected long-term behaviours (Figure \ref{fig:Eco_main_figure}C, green; Figure \ref{fig:ecol stability distribution}). However, both sources of information individually fail to capture important aspects of the population dynamics. Using the equilibrium constraints alone fails to capture population sizes and trends clearly seen from the dataset in Figure \ref{fig:Eco_main_figure}B (blue). Similarly, using the data alone means that there is no knowledge placed into the model indicating how long-term and stable these population fluctuations are, such that the data leads to systems with negative equilibrium abundances (Figure \ref{fig:Eco_main_figure}C, purple) and instability (Figure \ref{fig:ecol stability distribution}).

\subsubsection{Non-empirical information can alter forecasted ecosystem responses to management}

\begin{figure}[H]
    \centering
    \includegraphics[width=0.7\linewidth]{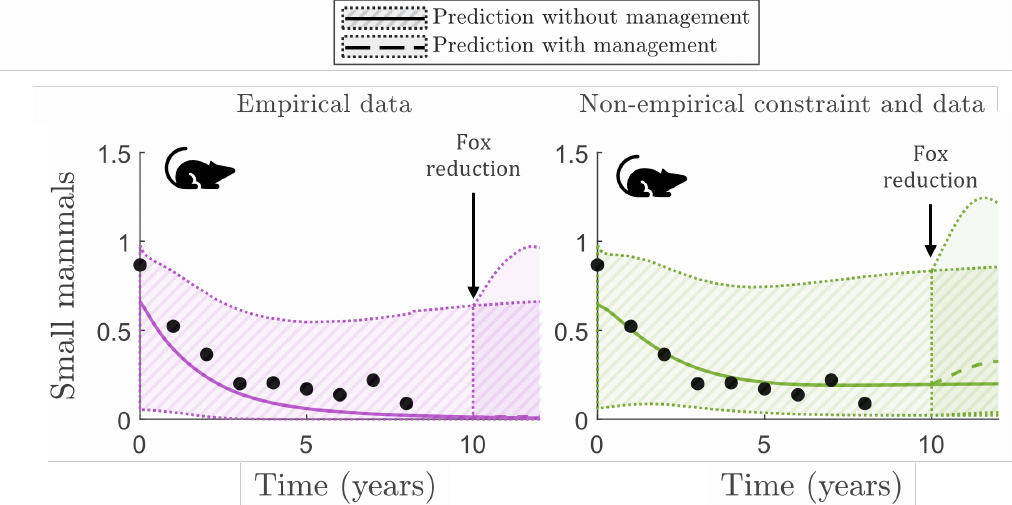}
    \caption{Predicted populations of small mammals, with and without regulating fox populations, across two parameter distributions: the distribution constrained by data (purple), and the distribution constrained by coexistence, stability and data (green). Notice that the constraints on coexistence and stability are necessary for reducing predictions that lead to small mammal extinction. }
    \label{fig:Ecol_management}
\end{figure}

Incorporating coexistence and stability requirements can cause fundamental differences in the long-term behaviour of the system, and can have major implications for ecosystem management. For a final ecosystem network scenario, we model the effects of reducing fox populations to 20\% of their predicted population after 10 years, and analyse the consequences for small mammal populations. Fitting a model to time-series data, without coexistence or stability requirements, yields predictions of the small mammal population heading towards extinction without fox control, and that heavily regulating the predator species, foxes, will also have minimal effect on small mammals (Figure \ref{fig:Ecol_management}, purple). 

Contrastingly, also fitting the model to coexistence and stability requirements informs the model that all four species' populations in the network are long-lasting despite their fluctuations. As a surprising consequence, the effects of regulating foxes are much more pronounced and suggest that small mammal populations will recover more strongly in response to this management action (Figure \ref{fig:Ecol_management}, green). Interestingly, in this case, fitting the model to both the non-empirical constraints and the dataset leads to a better match between the median of the model-data fit and the time-series data for the small mammal's population (Figure \ref{fig:Ecol_management}, green), compared to a model fit to the time-series data alone (Figure \ref{fig:Ecol_management}, purple). Whilst we do not expect this improvement in model-data fit due to other constraints to hold in general, it is additional evidence that non-empirical information can have pronounced and unexpected consequences on model predictions. 

\subsection{Case study 3: Biochemical adaptation}
\label{cellular adaptation case study}

A large body of literature utilises mathematical data-free techniques -- topological, analytical and numerical methods -- to identify biochemical systems that are capable of regulating and adapting to change \citep{ma_2009,jeynes_2023_protein_protein,araujo_2018_natcomms,khammash2021perfect,ferrell2016perfect}. Biochemical adaptation, or homeostasis, is a fundamental biological feature in which a system is able to repeatedly adapt to changes in a stimulus \citep{ma_2009,araujo_2018_natcomms,araujo_2023_PLOSCB,khammash2021perfect}; see Figure \ref{fig:Cellular main}B). However, analytical and topological approaches require strict assumptions yielding mathematical idealisations that are unlikely to occur in real systems \citep{ferrell2016perfect,aoki2019universal,briat2016antithetic}, such as requiring a system to perfectly return to its baseline, and allowing adaptation to occur on infinite timescales \citep{araujo_2018_natcomms,araujo_2023_PLOSCB,khammash2021perfect, karp2012complex}. Hence, numerical methods provide the best path for realistically representing the process of homeostasis in chemical reaction networks, where data is unavailable. 

In the field of biochemical adaptation, accept-reject methods for identifying homeostasis face a computational bottleneck that limits the size and consequently representativeness of the obtained ensemble of adaptation-capable parameter sets; for example, the seminal work of \cite{ma_2009} found that 1 in 10000 (0.01\%) tested parameterisations were capable of adaptation (\cite{jeynes_2023_protein_protein} and \cite{skataric2012characterization} report similar rates). To account for this problem, extended analyses of adaptation have often relied on reducing the parameter space being considered \citep{ma_2009} and separately manipulating subsets of parameters \citep{jeynes_2023_protein_protein}. However, these studies struggle to obtain a large and representative sample of parameter sets that lead to biochemical adaptation. 

An example of biochemical adaptation is seen in chemical reaction networks, such as the one depicted in Figure \ref{fig:Cellular main}A, that consist of a system of interacting proteins ($A,B$), enzymes ($E_1, E_4$) and complexes ($C_1,C_2,C_3,C_4$), typically represented as a system of ordinary differential equations \citep{ma_2009,jeynes_2023_protein_protein}. For this case study, we use a general modelling framework known as `complex complete' \citep{jeynes_2023_protein_protein} which explicitly accounts for the intermediate formation of complexes (the combination of enzymes and substrates) at the expense of additional parameters. This model consists of ten ordinary differential equations, and twelve highly uncertain reaction rate parameters that are typically drawn from independent log-uniform distributions spanning several orders of magnitude (\cite{jeynes_2023_protein_protein}; see Supplementary Materials Section \ref{SM:biochemical equations} for more detail). Under some specific parameter combinations, the network depicted in Figure \ref{fig:Cellular main}A can adapt to changes in input. To simulate this model for any given parameter set, the steady state is first numerically obtained, the input is changed (a process called ``stimulation''), and the model is solved under these new conditions to assess whether it returns to the same steady state or not (Figure \ref{fig:Cellular main}B).  

The system is considered adaptive if it meets two conditions after the input is changed: it must be sufficiently \textit{sensitive} and sufficiently \textit{precise} \citep{ma_2009}. Sensitivity, $S$, measures the ability of a chemical concentration, $O$, to react to a change in input, $I$,
\begin{align}
    S &= \frac{\left| O_{\rm{peak}}-O_1 \right| / O_1}{|I_2 - I_1|/I_1}, 
\end{align}
where $O_1$ and $I_1$ are the original chemical concentrations of the output and input at steady-state, respectively, $I_2$ is the updated input after stimulation, and $O_{\rm{peak}}$ is the concentration of $O$ which is furthest from $O_1$ after stimulation (see Figure \ref{fig:Cellular main}B). A biochemical network that is sufficiently sensitive to changes in input will have $S>1$ \citep{ma_2009}. 

The precision, $P$, is a measure of how well the system can return to its original value,
\begin{align}
    P &= \left( \frac{|O_2-O_1|/O_1}{|I_2-I_1|/I_1}\right)^{-1}, 
\end{align}
where $O_2$ is the final output after the input changes. A biochemical network is considered precise if $P>10$ \citep{ma_2009}. Hence, we can utilise these two summary statistics (sensitivity and precision) to define a measure of discrepancy $\rho$ between adaptive biochemical networks and a model simulation as
\begin{align}
    \rho &=  \rm{max}(0,1-\mathit{S}) + \rm{max}(0,10-\mathit{P}), 
\end{align}
such that $\rho$ measures the discrepancy from a sufficiently sensitive and precise system. Further details on the model, prior distribution, summary statistics, discrepancy, and likelihood can be found in the Supplementary Material Section \ref{SM:biochemical equations}. 

\subsubsection{Models can still be calibrated even without empirical data}

\begin{figure}[h!]
    \centering
    \includegraphics[width=\linewidth]{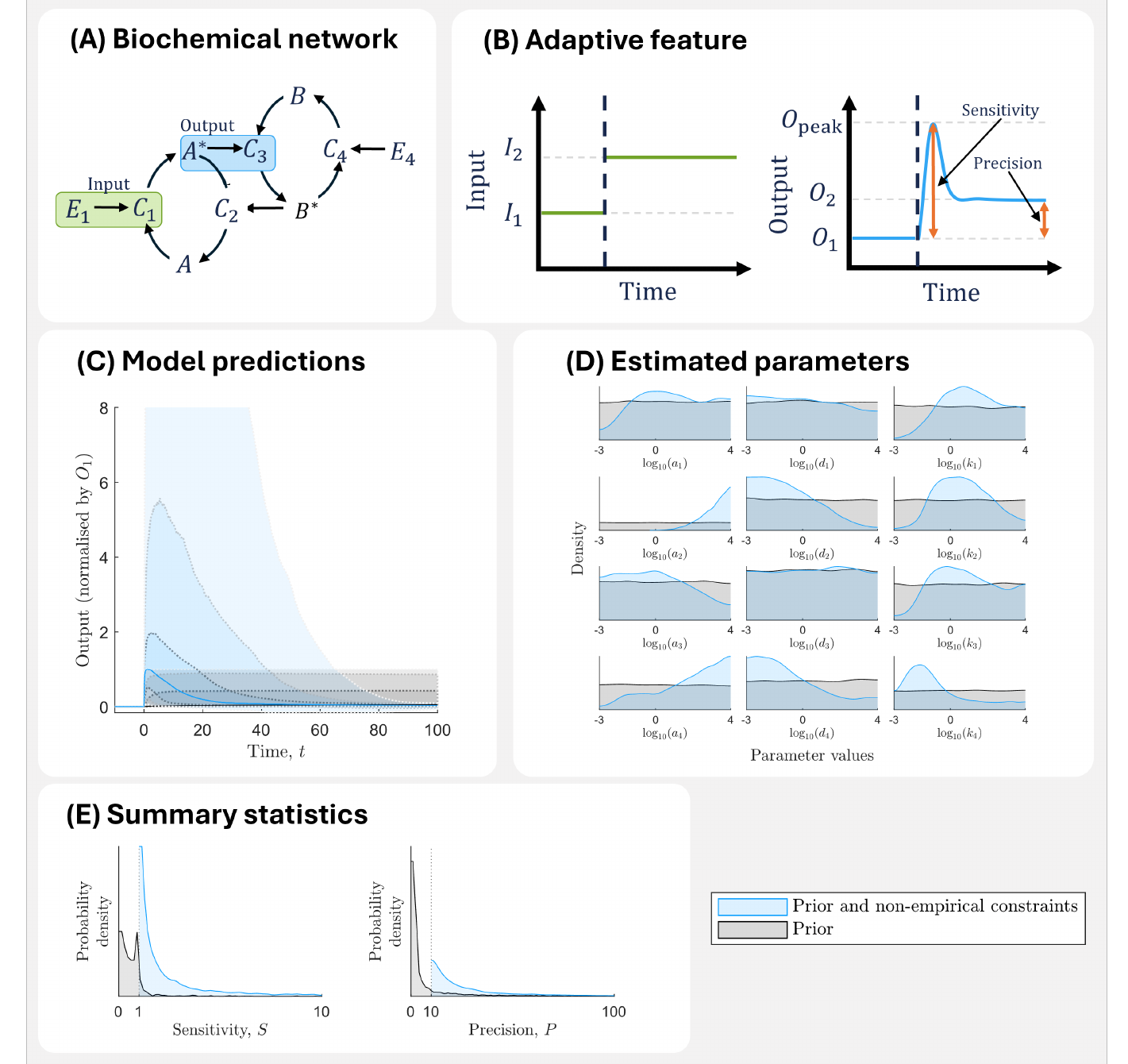}
    \caption{\small\textbf{(A)} A conceptual diagram of the biochemical reaction network capable of adaptation. This diagram depicts a system of protein substrates ($A$ and $B$), their activated forms ($A^*$ and $B^*$), enzymes ($E_1$ and $E_4$), and intermediate protein-protein complexes ($C_1, C_2, C_3$ and $C_4$) each interacting through biochemical reactions as illustrated by arrows. This biochemical network was used to construct the ordinary differential equation model. Note, in this example we are interested in the input $I=E_1 + C_1$ and the output $O=A^*+C_3$. \textbf{(B)} Conceptual diagram of biochemical adaptation, showing the two summary statistics: sensitivity and precision. As the levels of input change (from $I_1$ to $I_2$; left figure), the output is \textit{sensitive} to the change in stimulus, such that it changes from $O_1$ to a different concentration $O_\mathrm{peak}$. Additionally, the output is capable of \textit{precisely} returning to its pre-stimulated concentration, such that the output stabilises at a concentration $O_2$ that is similar to the pre-stimulus level $O_1$. \textbf{(C)} Model predictions from the prior distribution (grey) and the posterior distribution with the adaptive capability (blue). This figure shows the median prediction and a set of credible intervals (50\%, 75\% and 90\%) for the normalised output after the input has been perturbed at the initial time ($t=0$). \textbf{(D)} The estimated marginal distributions of model parameters both from the prior distribution (grey) and for model parameters that lead to sensitive and precise models (blue). This figure shows that there may be certain areas of parameter space that cannot yield a model capable of biochemical adaptation. \textbf{(E)} The measured sensitivity and precision of simulations both from the prior distribution (grey) and from the posterior distribution (blue), where model parameters were sufficiently sensitive ($S>1$) and precise ($P>10$) to be considered adaptive.}
    \label{fig:Cellular main}
\end{figure}

Using SMC-ABC with these summary statistics and discrepancy function, we produced an ensemble of 10,000 parameter sets that led to adaptation in this network. In comparison to simulations from the prior distribution, simulations from the posterior show the output is both sensitive to stimulation and precisely able to return to pre-stimulation levels (see Figure \ref{fig:Cellular main}C, Figure \ref{fig:Cellular main}E)).

\subsubsection{Deeper insight can be gained by using statistically robust methods to thoroughly search the parameter space}

\begin{figure}[h!]
    \centering
    \includegraphics[width=0.7\linewidth]{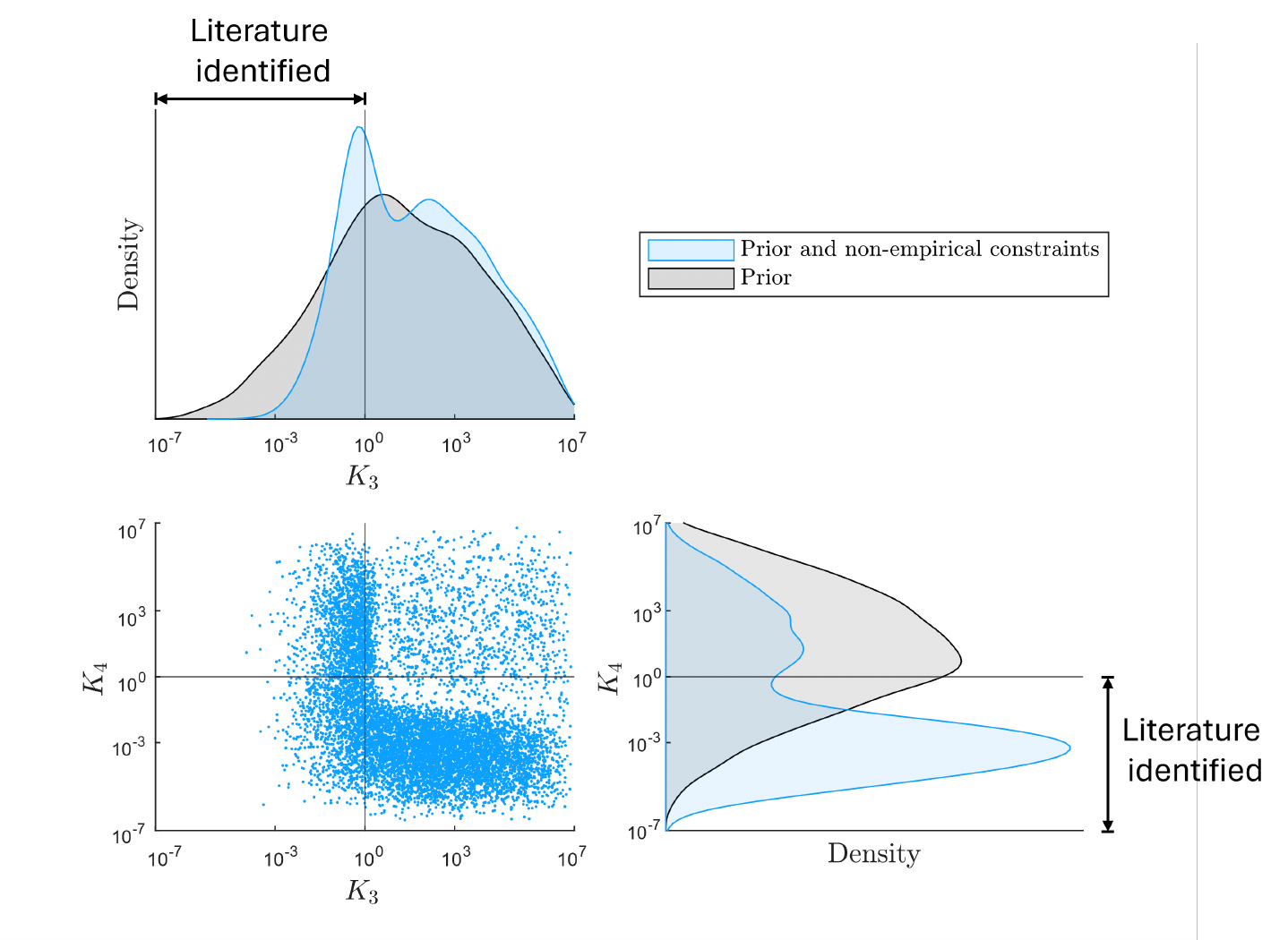}
    \caption{Distributions of the Michaelis-Menten constants $K_3$ and $K_4$ from the prior (grey), and the posterior (blue), where the posterior distribution represents models capable of biochemical adaptation, as well as the bivariate distribution from the posterior (blue dots). For comparison, the literature identified $K_3 <10^0$ and $K_4 < 10^0$ as necessary conditions for biochemical adaptation in this network \citep{jeynes_2023_protein_protein,ma_2009}.}
    \label{fig:cellular K3vsK4}
\end{figure}

An ensemble of parameter sets capable of adaptation (see Figure \ref{fig:Cellular main}D for parameter distributions) can be used to test interventions or analyse the network conditions that lead to the adaptive capability \citep{jeynes_2023_protein_protein}. Using a simplified modelling framework applied to the network in Figure \ref{fig:Cellular main}A, \cite{ma_2009} analytically showed that adaptation could only be achieved when the parameter combinations $K_3 = \frac{d_3+k_3}{a_3}$ and $K_4 = \frac{d_4+k_4}{a_4}$ were significantly smaller than 1.  While we generally observe this result within our parameter sets (Figure \ref{fig:cellular K3vsK4}, univariate distributions), our results numerically show that this condition is not strictly necessary for adaptation, and the bivariate distribution of these quantities of interest suggests that there may be some relationship between $K_3$ and $K_4$ (Figure \ref{fig:cellular K3vsK4}, bivariate distribution). 

This relationship may not have previously been uncovered because previous parameter searches have been more coarse (testing fewer parameter sets in the space) and have considered a smaller parameter space due to the low probability of identifying adaptation and the computational challenges of simulating these systems. For example, \cite{jeynes_2023_protein_protein} limited their search to a few orders of magnitude and held select parameters constant to explore the effects of stimulation range. Parameter sets that fall outside the area identified by the literature still display adaptive trajectories and do not appear to be different from those in the region where the literature suggests that adaptation is restricted to (Figure \ref{fig:cellular_K_group_comparison}). By switching to a robust and efficient sampling method, we have gained a deeper understanding of the quantities of interest in this model. Since these parameter combinations are related to enzymes and their binding affinity, which is highly relevant for drug development \citep{vellard_2003_enzyme_drugs,srinivasan_2023_enzyme_drugs}, effective mathematical modelling of these networks can offer medical insight and support.

\section{Discussion}
\subsection{Non-empirical information is powerful for calibration}
This work presents a general methodology to incorporate any information type -- whether empirical or non-empirical -- into model calibration. 
We demonstrate this methodology using three compelling examples from diverse scientific fields that illustrate how integrating non-empirical information can lead to better predictions, inference, management and understanding. 

Firstly, using a case study on coral colony recovery, we demonstrated how an expert’s knowledge of typical system function can be used in calibration to remove improbable model dynamics (Figure \ref{fig:logistic_constraints_only}). After calibration, the result is a model that both produces an excellent match to a dataset and aligns with expert beliefs of possible reef recovery dynamics (Figure \ref{fig:Logistic_predictions_data}). Given that applied models are often constructed in partnership with the system's stakeholders \citep{jakeman_2006,parrott_2017}, being able to query an expert’s knowledge of the system outputs and easily incorporate these into calibration is a huge asset, especially if done as an iterative process. For example, experts in coral reef recovery dynamics may have specific knowledge of reef recovery dynamics for highly damaged reefs \citep{Warne_2024, Warne_2022}, different types of disturbances \citep{graham_2011}, or for specific reef locations \citep{graham_2011}. Our results demonstrate that stakeholder knowledge can transform the predictive capabilities of a fitted model to possess drastically improved precision (Figure \ref{fig:logistic_constraints_only}B), whilst simultaneously refining plausible ranges on model parameters (Figure \ref{fig:Logistic_marginals}). 

Secondly, we generated an ensemble of ecosystem population models that match both short-term monitoring data and long-term behaviours that align with ecological theory (Figure \ref{fig:Eco_main_figure}). We demonstrated that using both information sources (monitoring data as empirical data and ecological theory as non-empirical information) can alter the predicted consequences of ecosystem management (Figure \ref{fig:Ecol_management}). Given the high uncertainty in parameter values of ecosystem network models \citep{baker_2019,vollert_2023_SMCEEM}, the limited availability of population time-series datasets \citep{humbert_2009,mcdonald_2010}, which otherwise yield unconstrained and uncertain model predictions \citep{botelho_2024,Novak_2011}, ecosystem network models stand to gain a lot from non-traditional data sources. Beyond the expected theoretical properties of ecosystems (coexistence and stability), there are several additional non-empirical information sources that can be used to inform ecosystem population models, including qualitative population responses to conservation (e.g.\ the observation that removing feral cats can lead to an increase in rabbit populations and consequently a decrease in native vegetation; see \cite{raymond_2011_unintendedConseq}), checking whether key model outputs are sensible (e.g., removing vegetation should result in herbivores like rabbits going extinct; see \cite{neil_2025}), or constraining other key quantities (e.g., restricting equilibrium or transient populations to reasonable domains; see \cite{vollert_2024_alternativeConstraints}). When constraining model predictions to fit these non-empirical expectations, the resulting model is better tied to reality, making it far more suitable for the conservation decision-making context it informs. 

Finally, we calibrated a model of a biochemical network to exhibit the observed phenomenon of biochemical adaptation (Figure \ref{fig:Cellular main}). In doing so, we were able to test and challenge conditions stated in the literature that are believed to be required for a biochemical network to express adaptation (Figure \ref{fig:cellular K3vsK4}). In future work, our approach could also be extended to explore and understand many other medically relevant features of homeostasis. For example, the discrepancy function used in our calibration method can include the sensitivity and precision to multiple stimuli to identify parameter sets capable of adaptation for a large range of inputs (see, e.g.\ \cite{jeynes_2023_protein_protein}); we can explore imperfect adaptation by reducing the requirement on precision tolerance (see, e.g.\ \cite{Bhattacharya_2023}); the time to adaptation can be incorporated to ensure that adaptation occurs in a biologically meaningful time-frame; and the model can be made stochastic to account for the randomness inherent in biochemical systems \citep{briat2016antithetic}. Prior to our study, it was not possible for these important concepts to be thoroughly explored via simulation due to computational limitations. Using our efficient and rigorous search algorithm, we generated a larger sample of parameter sets capable of biochemical adaptation than previous literature; e.g., \cite{ma_2009} and \cite{jeynes_2023_protein_protein} test $10^5$ parameter sets, with $\sim~1\%$ of these producing adaptation, whereas we have obtained $10^5$ parameter sets all of which are capable of adaptation. 

Across all three demonstrated examples, non-empirical data –- field knowledge of possible versus impossible system properties, theoretical long-term behaviour, and frequently observed phenomena –- were used to refine the plausible parameter space of simulation models, providing more realistic predictions (Figures \ref{fig:logistic_constraints_only}, \ref{fig:Logistic_predictions_data}, \ref{fig:Eco_main_figure} and \ref{fig:Cellular main}) and more informed inferences (Figures \ref{fig:Logistic_marginals}, \ref{fig:Logistic_marginals_data}, \ref{fig:ecol marginal} and \ref{fig:Cellular main}). For each case study, the results and findings are substantial for that field, yet we have only scratched the surface of what these methods could achieve. 

\subsection{Opportunities are unlocked by rigorous and systematic statistics}

The approach that we propose has three distinct methodological benefits: it provides a framework for calibration with a likelihood and approximate likelihood function simultaneously, it robustly and efficiently samples this distribution via our new SMC sampler, and it can be used as a novel simulation-based prior elicitation method.  

The literature for parameterising models using data is extensive \citep{Sisson_2018_ABC,mcelreath_2018,martin_2020}, but very few studies have looked at non-empirical constraints \citep{barnes_2011,vollert_2024_alternativeConstraints}. To the authors' knowledge, this is the first formal calibration framework for both data and non-empirical features simultaneously. Here, we have defined a distribution that combines a likelihood with a discrepancy function (Equation \eqref{eq:exact and approximate posterior}), allowing any information to be easily encoded into parameterisation via specification of these functions (see Methods section \ref{new methods}). This novel combination of data sources allows models to leverage the specificity of data, as well as the broad effects of non-empirical constraints. For example, in Section \ref{ecology case study}, \nameref{ecology case study}, we showed that population time-series datasets honed in on the magnitude of species populations (Figure \ref{fig:Eco_main_figure}B), whilst the constraints on long-term population dynamics ensured that the system as a whole functioned according to ecological theory (Figure \ref{fig:Eco_main_figure}C). While some models have been previously parameterised with a combination of data and non-empirical sources, these sampling algorithms have been ad-hoc, inefficient and not easily generalisable to new information (see e.g., typical processes in ecological network modelling \citep{neil_2025,Peterson_2021_DirkHartog,baker_2019} that rely on rejection sampling and optimisation algorithms). However, as a direct consequence of formalising the target distribution for calibration to non-empirical information (Equation \eqref{eq:exact and approximate posterior}), established sampling regimes can be leveraged. We developed an efficient and robust SMC algorithm that simultaneously anneals the likelihood and discrepancy (see Supplementary Material Section \ref{SM: sampler}).

Additionally, our statistical framework addresses a key challenge in Bayesian model calibration: defining the prior distribution \citep{Banner_2020}. Since many model parameters lack physical meaning or cannot be measured (e.g., consider the species interaction strengths in Section \ref{ecology case study}, \nameref{ecology case study}), developing methods for specifying prior distributions is an open challenge \citep{mikkola_2024}. Our method can be considered a novel approach to prior elicitation because it uses prior knowledge of reasonable outputs to infer an informed prior distribution via simulation. For example, in Section \ref{logistic case study}, \nameref{logistic case study}, we infer the informed prior distribution that yields plausible coral reef recovery dynamics according to prior knowledge of coral reef recovery times (Figure \ref{fig:Logistic_marginals}), and this informed prior can be simultaneously included in Bayesian inference with data (Figure \ref{fig:Logistic_marginals_data}). 

When considered as a prior elicitation method, our approach has three key advantages. Firstl, our method systematically includes prior predictive checks within our inference process, as all coral reef recovery predictions from the informed prior distribution must satisfy the expert's prior knowledge of recovery dynamics (Figure \ref{fig:logistic_constraints_only}). Second, in comparison to other prior elicitation methods, our approach is highly flexible and generalisable since it does not require manual specification of prior distributions to test (see e.g., \cite{wesner_2021}), or hyperparameters of the prior distribution (see e.g., \cite{bockting_2023}) and, therefore, does not rely on specifying a parametric family for the prior. Third, our prior elicitation method automatically captures the parameter interdependency structure via simulation, which is an otherwise ongoing challenge in prior elicitation \citep{mikkola_2024}, and resolving this problem is particularly important when constraints do not have a straightforward effect on individual parameter values (Figure \ref{fig:logistic_bivariate}). Hence, this new method represents a novel and generalisable simulation-based method for eliciting the joint prior distribution including its dependency structure.

\subsection{Building trust via model calibration in high uncertainty}
Whilst our methodology treats information for model calibration as generally irrefutable, we recognise that elicited knowledge can contain uncertainties and biases \citep{ohagan_2019}, which may need to be treated with uncertainty \citep{krueger_2012}, much like a likelihood function treats data with uncertainty. As such, confidence needs to be attributed to any non-empirical information, and practitioners should consider using conservative estimates, following elicitation guidelines \citep{hemming_2018}, and testing parameter sensitivity to non-empirical features \citep{saltelli_2006}. When the alternative is no data or limited data, it is our view that uncertain expert-elicited knowledge is still valuable and should not be ignored in model calibration. After all, empirical data can also be highly uncertain and biased, sparking concerns of trust in models built on such data \citep{harper_2021,vilas_2023}. However, through connecting to additional and varied sources of information, our approach may help to alleviate such concerns, e.g., by eliminating implausible model predictions (\nameref{logistic case study}), incorporating theory-based expectations (\nameref{ecology case study}), and broadening the parameter space when identifying models that recover observed system behaviours (\nameref{cellular adaptation case study}).

In this work, we have shown how non-empirical information can be mathematically utilised to provide a critically valuable resource for model calibration. Ultimately, leveraging the vast knowledge bases found in well-studied systems could be the key to better management and a better understanding of systems across science.

\section{Methods}
\subsection{Bayesian model calibration}
Model calibration (parameter estimation) is the process by which the parameters of a mathematical model are adjusted to align its outputs with the underlying observed system. Consider a model $M(\bm{\theta})$, where $\bm{\theta}$ is the set of model parameters. The goal of model calibration is to infer the parameters $\bm{\theta}$, such that the outputs of the model $M(\bm{\theta})$ are consistent with the data observed $\bm{y}_\mathrm{obs}$. In Bayesian model calibration we aim to obtain a probability distribution of values $\bm{\theta}$ according to both the likelihood of the data arising from parameter values and the prior probability of parameter values. Bayesian inference provides a coherent statistical framework for combining this information to form a posterior distribution \citep{mcelreath_2018,Girolami_2008,martin_2020}: 
\begin{equation}
\label{eq:exact posterior}
    \pi(\bm{\theta}|\bm{y}_\mathrm{obs}) \propto \pi(\bm{\theta}) f(\bm{y}_\mathrm{obs}|\bm{\theta}).
\end{equation}
Here, $\pi(\bm{\theta})$ is the prior distribution, representing our initial belief about the parameters $\bm{\theta}$ before observing any data, and $f(\bm{y}_\mathrm{obs}|\bm{\theta})$ is the likelihood function, representing the probability of the observed data $\bm{y}_\mathrm{obs}$ given the parameters $\bm{\theta}$ when generated using the model $M(\bm{\theta})$. The posterior distribution $\pi(\bm{\theta}|\bm{y}_\mathrm{obs})$ combines the prior information with the data to yield a representation of the probability of model parameter values $\bm{\theta}$ for the given observed data $\bm{y}_\mathrm{obs}$.

Bayesian model calibration is a foundational tool for finding an appropriate distribution of parameter values that match mathematical models to the data. Using a sampling algorithm -- such as Markov chain Monte Carlo (MCMC; \citeauthor{gamerman_2006}, \citeyear{gamerman_2006}) or sequential Monte Carlo (SMC; \citeauthor{delmoral_2006}, \citeyear{delmoral_2006}; \citeauthor{chopin_2002}, \citeyear{chopin_2002}) -- an ensemble of parameter samples from the posterior distribution can be obtained and fed into these models for predictions that aim to match real-world observations. However, it remains a challenge to understand how non-empirical data fits within this framework because the likelihood function cannot be easily defined.

\subsection{Approximate Bayesian inference}
Approximate Bayesian computation (ABC) is a model calibration framework used when the likelihood cannot be calculated. These methods approximate the likelihood function by comparing model simulations to observations and measuring the discrepancy between them \citep{Beaumont_2019_ABC,Sisson_2018_ABC,sunnaaker_2013_ABC,Beaumont_2010_ABC}. Here, we consider ABC methods because the likelihood may not be calculable when the data source is non-empirical, such as expert judgment or qualitative insights. For example, it would be challenging to define the likelihood function for an ecologist's observation that ecosystem populations must have a stable equilibrium (see Section \ref{ecology case study}, \nameref{ecology case study}). 

Approximate Bayesian frameworks compare simulated data ${\bm{y}_\mathrm{sim}}(\bm{\theta})$ for the model specified by  $\bm{\theta}$, to the observed data ${\bm{y}_\mathrm{obs}}$ using two key functions: a summarisation function, and a discrepancy function. The summarisation function $S(\cdot)$ reduces the information from the full dataset $\bm{y}_\mathrm{obs}$ or model simulation $\bm{y}_\mathrm{sim}(\bm{\theta})$ to a small set of comparable summary statistics $\bm{S}_\mathrm{obs} = S(\bm{y}_\mathrm{obs})$ and $\bm{S}_\mathrm{sim}(\bm{\theta}) = S(\bm{y}_\mathrm{sim}(\bm{\theta}))$ which retain the key information needed to compare $\bm{y}_\mathrm{obs}$ to ${\bm{y}_\mathrm{sim}}(\bm{\theta})$; e.g., the mean and variance are common summary statistics. A discrepancy function $\rho(\cdot)$ is then used to characterise the difference between the summary statistics for the observed data $\bm{S}_\mathrm{obs}$ and the simulated data $\bm{S}_\mathrm{sim}(\bm{\theta})$. As such, the discrepancy  $\rho \equiv \rho(\bm{S}_\mathrm{obs}, \bm{S}_\mathrm{sim}(\bm{\theta}))$ can be used to assess whether the parameter values are plausible based on how closely the summary statistics match. This discrepancy function is non-negative and produces a scalar discrepancy score $\rho$ that is $0$ if $\bm{S}_\mathrm{obs} = \bm{S}_\mathrm{sim}(\bm{\theta})$ and $>0$ when $\bm{S}_\mathrm{obs} \neq \bm{S}_\mathrm{sim}(\bm{\theta})$.

In ABC, a parameter set $\bm{\theta}$ is typically accepted as plausible if the discrepancy $\rho$ between simulated and observed summaries is below some threshold $\epsilon$. Therefore, the discrepancy function can be used to approximate the likelihood, as both the discrepancy and likelihood assess the ability of the parameters to produce the observed outputs:
\begin{align}
\nonumber
    f(\bm{y}_\mathrm{obs}|\bm{\theta}) & \approx  
    \int \mathbb{I}\Bigl( \rho \bigl( S(\bm{y}_\mathrm{sim}(\bm{\theta})), S(\bm{y}_\mathrm{obs})\bigr)< \epsilon \Big|\bm{\theta} \Bigr) \ \mathrm{d}\bm{y}_\mathrm{sim},
\end{align}
where $\mathbb{I}(\cdot)$ is an indicator function that is 1 when the condition inside is true, and 0 otherwise. Using this approximation of the likelihood, the posterior distribution $\pi(\bm{\theta}|\bm{y}_\mathrm{obs})$ can be approximated as $\pi_\epsilon(\bm{\theta}|\bm{y}_\mathrm{obs})$ given by 
\begin{align}
\nonumber
    \pi_\epsilon(\bm{\theta}|\bm{y}_\mathrm{obs}) &\propto  
    \pi(\bm{\theta}) \int \mathbb{I}\Bigl( \rho \bigl(S(\bm{y}_\mathrm{sim}(\bm{\theta})), S(\bm{y}_\mathrm{obs})\bigr)< \epsilon \Big|\bm{\theta} \Bigr)  \ \mathrm{d}\bm{y}_\mathrm{sim}.
\end{align} 
Parameter values from the approximate posterior $\pi_\epsilon(\bm{\theta}|\bm{y}_\mathrm{obs})$ can be sampled using a variety of methods, such as ABC-MCMC \citep{marjoram2003markov} or ABC-SMC \citep{delmoral_2012}; see \cite{Sisson_2018_ABC} or \cite{Beaumont_2019_ABC} for an overview.

\subsection{Incorporating non-empirical data in Bayesian inference}
\label{new methods}
Traditionally, ABC converts empirical data to carefully chosen summary statistics for comparison. However, if the summarised features of the data can be obtained directly -- for example, via expert elicitation, observation, logic, physical rules, or even hypothesis -- these could also be used for model calibration. In traditional ABC, summary statistics are calculated from empirical data as $\bm{S}_\mathrm{obs} = S(\bm{y}_\mathrm{obs})$, here we relax this requirement such that non-empirical data sources are treated as directly observed summary statistics $\bm{S}_\mathrm{obs}$. For example, an ecologist's observation that coral populations can recover to carrying capacity within 50 years can be used as a summary statistic even if there is no empirical data available to estimate this statistic (see Section \ref{logistic case study}, \nameref{logistic case study}). In our non-empirical framework, the summary statistics are constructed based on (a) what information $\bm{S}_\mathrm{obs}$ can be known with confidence, and (b) whether this summary statistic can be measured from model outputs $\bm{S}_\mathrm{sim} = \bm{S}(\bm{y}_\mathrm{sim})$. For example, an ecologist must be confident that coral cover cannot exceed 10\% within 5 years post-disturbance, and the model must be able to simulate coral cover through time (see Section \ref{logistic case study}, \nameref{logistic case study}).

Compared to summary statistics calculated from empirical data, we often cannot elicit an exact value of a summary statistic from non-empirical sources. Continuing with our coral cover example, if data were routinely collected at various reefs 5 years after a disturbance, then the mean coral cover 5 years post-disturbance could be calculated from the empirical dataset. However, in the absence of this empirical data, it is a challenge for an ecologist to estimate the mean coral cover with confidence, and it is far easier to elicit a range of values for the summary statistic, such that coral cover should be less than 10\%.

Whether the summary statistic is an exact value (calculated from an empirical dataset, $S_\mathrm{obs} = S(\bm{y}_\mathrm{obs})=a$) or a range of values (elicited from experts, $S_\mathrm{obs} \leq a$) is an artefact of the available data for model calibration, yet having a range of values for a summary statistic is a relatively strange concept in ABC that requires subtle changes to the statistical formulation (see Table \ref{tab:traditional_vs_nonempirical_ABC}). A traditional ABC discrepancy function $\rho(\cdot)$ would describe the distance between the observed and simulated summary statistic, and this discrepancy score would be minimised to some tolerance $\epsilon$ to approximate the likelihood. Instead, where the observed summary statistic is expressed as a range, the discrepancy must measure the distance to the acceptable range (achieved using minimisation or maximisation functions) and we aim to find parameter sets that lead to no discrepancy $\epsilon=0$. In the coral growth example, this discrepancy would measure how far above 10\% the coral cover is after 5 years, and we would only accept models with no discrepancy, i.e., no predictions over 10\%. These subtle differences from empirical ABC (summarised in Table \ref{tab:traditional_vs_nonempirical_ABC}) shift the calibration towards rejecting models with impossible outputs, rather than closely matching empirical datasets. 

\begin{table}[H]
    \centering
    \begin{tabular}{lcc}
        & \textbf{Empirical ABC} & \textbf{Non-empirical ABC}\\ 
        Observed summary statistic & $S_\mathrm{obs} = S(\bm{y}_\mathrm{obs})=a$ & $S_\mathrm{obs} \leq a$\\ 
        Simulated summary statistic & $S_\mathrm{sim}(\bm{\theta}) =  S(\bm{y}_\mathrm{sim}(\bm{\theta}))=b$ & $ S_\mathrm{sim}(\bm{\theta}) = S(\bm{y}_\mathrm{sim}(\bm{\theta}))=b$\\ 
        Discrepancy function & $ \rho(S_\mathrm{obs},S_\mathrm{sim})= |b-a|$ & $ \rho(S_\mathrm{obs},S_\mathrm{sim})= \mathrm{max}(b-a,0)$\\ 
        Likelihood approximation & $\mathbb{I}(\rho(S_\mathrm{obs},S_\mathrm{sim})<\epsilon)$ & $\mathbb{I}(\rho(S_\mathrm{obs},S_\mathrm{sim})=0)$
    \end{tabular}
    \caption{An example of the potential differences between empirical and non-empirical ABC.}
    \label{tab:traditional_vs_nonempirical_ABC}
\end{table}

For a given set of non-empirical data $\bm{S}_\mathrm{obs}$ and their corresponding summary statistics and discrepancy function (e.g., as in Table \ref{tab:traditional_vs_nonempirical_ABC}), an approximate posterior can be defined as
\begin{align}
\nonumber
    \pi(\bm{\theta}|\bm{S}_\mathrm{obs}) & \propto 
    \pi(\bm{\theta}) \int \mathbb{I}\Bigl( \rho \bigl(\bm{S}_\mathrm{obs},  S(\bm{y}_\mathrm{sim}(\bm{\theta}))\bigr)=0 \Big|\bm{\theta} \Bigr)  \ \mathrm{d}\bm{y}_\mathrm{sim}, 
    \label{eq:informed prior}
\end{align}
and parameter sets sampled using any ABC sampling approach. 

Additionally, non-empirical data can be combined with empirical data; via a likelihood function, such as a Gaussian likelihood, 
\begin{equation}
    f(\bm{y}_\mathrm{obs}|\bm{\theta}) = \prod _{i=1}^nf(y_{\mathrm{obs},i}|\bm{\theta}) = \prod _{i=1}^n \frac{1}{\sqrt{2\pi} \sigma} \exp \left( -\frac{y_{\mathrm{sim},i}(\bm{\theta}) - y_{\mathrm{obs},i})^2}{2\sigma^2} \right)
    \label{eq:gaussian}
\end{equation}
where $f(\bm{y}_\mathrm{obs}|\bm{\theta})$ is the likelihood of observing data $\bm{y}_\mathrm{obs}$ from a model characterised by parameters $\bm{\theta}$, $n$ is the number of observed data points, $y_{\mathrm{obs},i}$ is the $i$th observation, $\sigma$ is a parameter representing measurement noise, and $y_{\mathrm{sim},i}$ is the data simulated by the model characterised by parameters $\bm{\theta}$ with equivalent inputs to that of data point $i$. Using a likelihood to describe the probability that the model defined by a set of parameters $\bm{\theta}$ produced the dataset $\bm{y}_{\rm{obs}}$, we can target both the non-empirical observations and the traditional Bayesian posterior (as in Equation \eqref{eq:exact posterior}); 
\begin{align}
    \pi(\bm{\theta}|S_\mathrm{obs}, \bm{y}_\mathrm{obs}) &\propto \pi(\bm{\theta}) f(\bm{y}_\mathrm{obs}|\bm{\theta}) \int \mathbb{I}\Bigl( \rho \bigl(S_\mathrm{obs},  S(\bm{y}_\mathrm{sim}(\bm{\theta}))\bigr)=0 \Big|\bm{\theta} \Bigr)  \ \mathrm{d}\bm{y}_\mathrm{sim}.
    \label{eq:exact and approximate posterior} 
\end{align}
This combination of empirical and non-empirical data means highly data-limited scenarios can still take advantage of any available information. 

Sampling this distribution can be done by leveraging existing methods of sampling an exact or approximate distribution. In this work, we have developed a novel modification of the SMC-ABC algorithm proposed by \cite{drovandi_2011_ABC}, to efficiently sample the distribution in Equation \eqref{eq:exact and approximate posterior}. This algorithm sequentially reduces the discrepancy and anneals the likelihood by iteratively reweighting, resampling and moving an ensemble of parameter sets until a posterior distribution with no discrepancy is obtained (see Supplementary Material \ref{SM: sampler} for details). This adaptive and simultaneous incorporation of both the likelihood and approximate likelihood allows for flexible, efficient and robust sampling of the distribution which both matches the data and the non-empirical constraints. 

\subsection*{Data and code availability}
All data needed to evaluate the conclusions in the paper are present in the paper and/or the Supplementary Materials. MATLAB code used to generate the results in this paper is available in the Figshare Repository: \url{https://doi.org/10.6084/m9.figshare.29115044}. Code to use the methods developed in this paper is available in R (\url{https://github.com/luzvpascal/SMCfeatures}) and in MATLAB (\url{https://doi.org/10.6084/m9.figshare.29115044}). 

\subsection*{Acknowledgements}
SAV is supported by a Queensland University of Technology Centre for Data Science, Australia Scholarship and a Statistical Society of Australia Top-Up Scholarship. CD and LVP are supported by an Australian Research Council Discovery Project (DP200102101). LVP acknowledges funding from an Australian Research Council Discovery Early Career Researcher Award (DE200101791). MPA and SAV acknowledge funding support from an Australian Research Council Discovery Early Career Researcher Award (DE200100683). LVP and MPA acknowledge funding from the ARC SRIEAS Grant SR200100005 Securing Antarctica’s Environmental Future. The eResearch Office, Queensland University of Technology provided computational resources.

\newpage
\section*{References}
\begingroup
\renewcommand{\section}[2]{}%
\bibliographystyle{chicagoa}

\endgroup

\newpage
\appendix
\section*{Supplementary materials}
\setcounter{subsection}{0}
\setcounter{table}{0}
\setcounter{figure}{0}
\setcounter{equation}{0}
\setcounter{algocf}{0}
\renewcommand{\thesubsection}{S.\arabic{subsection}}
\renewcommand{\thetable}{S\arabic{table}}
\renewcommand{\thefigure}{S\arabic{figure}}
\renewcommand{\theequation}{S\arabic{equation}}
\renewcommand{\thealgocf}{S\arabic{algocf}}

\begingroup
\renewcommand{\section}[2]{}%

{\renewcommand{\baselinestretch}{1}} 

\subsection{Additional modelling details for each of the case studies}
This work demonstrates the value of using non-empirical information within model calibration for mechanistic models. To demonstrate our methods, we will use three different modelling examples: a case study of logistic coral growth calibrated with expert-elicited knowledge, a case study of ecosystem population modelling calibrated using ubiquitous theories of long-term population dynamics, and a case study of biochemical reaction networks calibrated to be adaptive to changes in input. In this section, we provide additional details of the modelling scenarios that may be necessary to replicate our study. 

\subsubsection{Case study 1: Logistic coral growth}
\label{SM: logistic details}

Section \ref{logistic case study}, \nameref{logistic case study} in the main text describes a simple modelling scenario, where the percentage of coral cover on a reef following a disturbance is described using a logistic model. In this example, the percentage of coral cover on a reef can be modelled as
\begin{equation}
    \frac{\rm{d}\mathit{y}}{\rm{d}\mathit{t}} = ry(t)\left(1-\frac{y(t)}{K}\right),\qquad y(0) = y_0,
\end{equation}
where $y$ is the coral cover (\% area), $t$ is the time (years), $r$ is a growth rate parameter (year$^{-1}$), $K$ is the carrying capacity (\% area), and $y_0$ is the initial coral cover (\% area). This ordinary differential equation can be solved analytically as 
\begin{equation}
    y(t) = \frac{K y_0}{y_0+(K-y_0)e^{-rt}}.
\end{equation}
This model has three model parameters to be calibrated (Table \ref{tab:logistic_parameters}). Since this is an illustrative example, the specified prior distributions are arbitrary, however, the bounds are loosely based on other works in the literature \citep{simpson_2022_sigmoidReefs,Simpson_2023}. 

\begin{table}[H]
    \centering
    \begin{tabular}{c|c|c|c}
        \textbf{Parameter} & \textbf{Description} & \textbf{Units} & \textbf{Prior distribution} \\ 
         $r$ & Growth rate & $1/$year & $U(0, 0.5)$ \\
         $K$ & Maximum coral cover & \% area & $U(60, 80)$ \\
         $y_0$ & initial coral cover & \% area & $U(0, 5)$ \\
    \end{tabular}
    \caption{The model parameters associated with the logistic growth model used in Section \ref{logistic case study}, \nameref{logistic case study}. }
    \label{tab:logistic_parameters}
\end{table}

Two non-empirical constraints were considered. Firstly, we constrain acceptable parameter sets by requiring slow initial growth, such that coral cover should be less than 10\% within the first 5 years. Secondly, acceptable parameter sets must exhibit recovery, such that coral populations should be within 1\% percentage area of carrying capacity within 50 years. Both of these statements correspond to a summary statistic, that is expected to be within some bounds. Here, these conditions can be expressed as
\begin{align}
    y(5) &= \frac{K y_0}{y_0+(K-y_0)e^{-5r}} \leq 10\%, \\
    y(50) &= \frac{K y_0}{y_0+(K-y_0)e^{-50r}} \geq K-1\%,
\end{align} 
such that $y(5)$ and $y(50)$ are summary statistics in the model. These summary statistics can then be easily compared to the expert knowledge via one discrepancy function $\rho$: 
\begin{align}
    \rho &= \rho_{5} + \rho_{50},    \\
    \rho_{5} &= \mathrm{max}(0, y(5) - 10), \\
    \rho_{50} &= \mathrm{max}(0, K -1 - y(50)),    
\end{align}
where $\rho_{5}$ measures the discrepancy of coral cover exceeding 10\% in the first 5 years, $\rho_{50}$ measures the discrepancy of coral cover not recovering within 1\% of $K$ in 50 years, and $\rho$ measures the total discrepancy between the simulated and expected coral cover.  

In addition, we generated three sparse observations of coral cover (Table \ref{tab:logistic_dataset}) to demonstrate how data can be used within our framework. 

\begin{table}[H]
    \centering
    \begin{tabular}{c|c}
        \textbf{Time (years)} & \textbf{Coral cover (\% area)} \\
        1 & 4\\
        3 & 4\\
        10 & 10
    \end{tabular}
    \caption{The time-series dataset used to calibrate the logistic growth model in Section \ref{logistic case study}, \nameref{logistic case study}. }
    \label{tab:logistic_dataset}
\end{table}

To calibrate the model to this dataset we used a Gaussian likelihood, which assumes normally distributed measurement noise. 
\begin{align}
    f(\bm{y}_\mathrm{obs}|\bm{\theta}) &= \prod _{i=1}^nf(y_{\mathrm{obs},i}|\bm{\theta}) \\
    &= \prod _{i=1}^n \frac{1}{\sqrt{2\pi} \sigma} \exp \left( -\frac{y_{\mathrm{sim},i}(\bm{\theta}) - y_{\mathrm{obs},i})^2}{2\sigma^2} \right)
    \label{eq:gaussian likelihood}
\end{align}
where $f(\bm{y}_\mathrm{obs}|\bm{\theta})$ is the likelihood of observing data $\bm{y}_\mathrm{obs}$ from a model characterised by parameters $\bm{\theta}$, $n$ is the number of observed data points, $y_{\mathrm{obs},i}$ is the $i$th observation, $\sigma$ is a parameter representing measurement noise, and $y_{\mathrm{sim},i}$ is the data simulated by the model characterised by parameters $\bm{\theta}$ with equivalent inputs to that of data point $i$. Hence, for the data described in Table \ref{tab:logistic_dataset}, $n=3$, $\bm{y}_\mathrm{obs}=\{ 4, 4, 10\}$, and $\bm{y}_{\mathrm{sim}} = \{ y(1), y(3), y(10)\}$.  

We generated 10000 parameter sets for each of the target distributions using our sequential Monte Carlo (SMC) algorithm that blends exact and approximate approaches. Details on this sampling algorithm can be found in Supplementary Material Section \ref{SM: sampler}.  

\subsubsection{Case study 2: Ecosystem population modelling}
\label{SM: ecosystem details}

Section \ref{ecology case study}, \nameref{ecology case study} in the main text describes an ecosystem population model, used to model the populations of four interacting species for making management decisions. In this paper, we focus on modelling the network depicted in Figure \ref{fig:Eco_main_figure}A, comprised of foxes ($F$), rabbits ($R$), small mammals ($M$) and vegetation ($V$). We use the generalised Lotka-Volterra equations to model this network as a system of ordinary differential equations, which is used to model population predictions in an ecosystem as
\begin{equation}
        \frac{\mathrm{d}n_i}{\mathrm{d}t}= \left[ r_i + \sum_{j=1}^{N} \alpha_{i,j} n_j(t) \right] n_i(t),
\end{equation}
where $n_i(t)$ is the abundance of the $i$th species at time $t$, $r_i$ is the growth rate of the $i$th species, $N$ is the number of species being modelled, and $\alpha_{i,j}$ is the per-capita interaction strength characterising the effect of species $j$ on species $i$.  The presence and sign of interactions between species $\alpha_{i,j}$ are prescribed by the ecosystem network (e.g., fox populations negatively affect rabbit populations therefore $\alpha_{R,F}<0$, but fox populations have no direct effect on vegetation abundance so $\alpha_{V,F}=0$). 

For the network we consider, the full Lotka-Volterra ecosystem model can be written as
\begin{align}
    \frac{\mathrm{d}n_F}{\mathrm{d}t} &= \left[ r_F + \alpha_{F,F} n_F(t)+ \alpha_{F,R} n_R(t)+ \alpha_{F,M} n_M(t)  \right] n_F(t), \\
    \frac{\mathrm{d}n_R}{\mathrm{d}t} &= \left[ r_R + \alpha_{R,F} n_F(t)+ \alpha_{R,R} n_R(t)+ \alpha_{R,V} n_V(t)  \right] n_R(t), \\
    \frac{\mathrm{d}n_M}{\mathrm{d}t} &= \left[ r_M + \alpha_{M,F} n_F(t)+ \alpha_{M,M} n_M(t)+ \alpha_{M,V} n_V(t)  \right] n_M(t), \\
    \frac{\mathrm{d}n_V}{\mathrm{d}t} &= \left[ r_V + \alpha_{V,R} n_R(t)+ \alpha_{V,M} n_M(t)+ \alpha_{V,V} n_V(t)  \right] n_V(t). 
\end{align}

Alternatively, we can express this model in vector form, such that 
\begin{equation}
    \frac{\mathrm{d}\bm{n}}{\mathrm{d}t}=[\bm{r} + \bm{A} \bm{n}] \circ \bm{n},
    \label{eq: matrix LV}
\end{equation}
where $\bm{n}=\{n_i:i=F,R,M,V\}$ is a vector of populations for each species, $\bm{r}=\{r_i:i=F,R,M,V\}$ is the vector of growth rates for each group, $\bm{A}=\{\alpha_{i,j}:i,j=F,R,M,V\}$ is the $N\times N$ interaction matrix of per-capita interaction strengths between ecosystem nodes, $\circ$ is the Hadamard or element-wise product, and the subscripts $F, R, M,$ and $V$ refer to foxes, rabbits, small mammals and vegetation, respectively. 

To solve this system of ordinary differential equations, we additionally require initial populations for each of the species $n_i(0)$ where $i=\{F,R,M,V\}$. As such, 16 parameters in this model require calibration (see Table \ref{tab:ecological_parameters}).

\begin{table}[H]
    \centering
    \begin{tabular}{p{0.35\linewidth}|p{0.4\linewidth}|p{0.15\linewidth}}
        \textbf{Parameter} & \textbf{Description} & \textbf{Prior} \\ 
         $n_F(0)$ & Initial fox population & $U(0.25,1)$\\
         $n_R(0)$ & Initial rabbit population & $U(0.5,1)$\\
         $n_M(0)$ & Initial small mammal population & $U(0,1)$\\
         $n_V(0)$ & Initial vegetation population & $U(0.25,1)$\\
         $r_F$, \ $r_R$, \ $r_M$, and $r_V$ & Species growth rates & $U(-1,1)$ \\
         $\alpha_{F,F}$, \ $\alpha_{R,R}$, \ $\alpha_{M,M}$, and $\alpha_{V,V}$  & Intra-species interaction & $U(-1,0)$\\
         $\alpha_{R,F}$, \ $\alpha_{M,F}$, \ $\alpha_{V,R}$, and $\alpha_{V,M}$ & Negative species interaction & $U(-1,0)$ \\
         $\alpha_{F,R}$, \ $\alpha_{F,M}$, \ $\alpha_{R,V}$, and $\alpha_{M,V}$ & Positive species interaction & $U(0,1)$ \\
    \end{tabular}
    \caption{The model parameters associated with the logistic growth model used in Section \ref{logistic case study}, \nameref{logistic case study}. }
    \label{tab:ecological_parameters}
\end{table}

In this ecosystem example, we consider two non-empirical constraints: equilibrium feasibility and stability. \textit{Feasibility} (often referred to as coexistence) specifically measures whether the equilibrium populations of each population are positive \citep{Grilli_2017_FS}, where the equilibrium populations are calculated as
\begin{align}
   \bm{n}^* &= -\bm{A}^{-1}\bm{r} > \bm{0},
   \label{eq: equilibrium matrix}
\end{align}
where $\bm{n}^*$ is the vector of equilibrium population abundances for all species $n^*_i = \{i:C,R,B,V\}$. These four equilibrium abundances are each summary statistics that must all be positive to meet the feasibility condition. 

\textit{Stability} measures the ability of the equilibrium to resist changes from small external pressures \citep{Allesina_2012_stab} and is measured by taking the eigenvalues $\bm{\lambda}$ of the Jacobian matrix $\bm{J}$ evaluated at equilibrium $\bm{n}^*$. For Lotka-Volterra each element $(i,j)$ of the Jacobian $\bm{J}$ is calculated as $J_{i,j} = \alpha_{i,j}n_i^*, \ \forall i,j$ in $\{F,R,M,V\}$ such that the summary statistics for stability are real components of the four eigenvalues
\begin{equation}
    \mathbb{R}(\bm{\lambda}) < \bm{0}
\end{equation}
of the Jacobian matrix $\bm{J}$ and must each be negative for a stable equilibrium. For a more detailed explanation of the calculation of these summary statistics see \cite{vollert_2023_SMCEEM}. All eight summary statistics (four equilibrium abundances $n_i^*$ for each species $i=\{F,R,M,V\}$, and four real components of eigenvalues $\mathbb{R}(\lambda_i)$ for each dimension in the population-space) are then combined within a single discrepancy, calculated as 
\begin{align}
    \rho &=  \sum_{i=1}^4 \big| \min \{ 0,n_i^*) \} \big|  + \sum_{i=1}^4 \big| \max\{0,\mathbb{R}(\lambda_i)\} \big|,
\end{align}
such that $\rho$ measures the discrepancy in equilibrium behaviour from the theorised feasible and stable behaviour. 

In addition to calibrating this model with non-empirical constraints, we calibrate the model to a simulated time-series dataset of nine observations per species (see Table \ref{tab:ecosystem_dataset}). 

\begin{table}[H]
    \centering
    \begin{tabular}{c|c|c|c|c}
        \textbf{Time} & \multicolumn{4}{c}{\textbf{Populations}}\\ \hline
         & \textbf{Fox} & \textbf{Rabbit} & \textbf{Small mammal} & \textbf{Vegetation}\\ 
        0 & 	0.4641	& 	0.6850	& 	0.8673	& 	0.7174	\\
        1 & 	0.5877	& 	0.7865	& 	0.5226	& 	0.6179	\\
        2 & 	1.0489	& 	1.0748	& 	0.3640	& 	0.4555	\\
        3 & 	1.4754	& 	0.8941	& 	0.1997	& 	0.5363	\\
        4 & 	1.7064	& 	0.7186	& 	0.2047	& 	0.4482	\\
        5 & 	1.6375	& 	0.7417	& 	0.1703	& 	0.6880	\\
        6 & 	1.4619	& 	0.6292	& 	0.1371	& 	0.8902	\\
        7 & 	1.6266	& 	0.6306	& 	0.2197	& 	1.1218	\\
        8 & 	1.4009	& 	0.8132	& 	0.0883	& 	1.1005	
    \end{tabular}
    \caption{The time-series dataset used to calibrate the ecosystem population model in Section \ref{ecology case study}, \nameref{ecology case study}. }
    \label{tab:ecosystem_dataset}
\end{table}

We used a Gaussian likelihood to calibrate the model to this dataset (see Equation \eqref{eq:gaussian}), which assumes normally distributed measurement noise. Specifically, for ecosystem population modelling, the likelihood function was defined as
\begin{align}
    f(\bm{y}_\mathrm{obs}|\bm{\theta}) = \prod _{i=1}^N \prod _{j=1}^D \frac{1}{\sqrt{2\pi} \sigma_i} \exp \left( -\frac{y_{\mathrm{sim},i}(t_j,\bm{\theta}) - y_{\mathrm{obs},i}(t_j))^2}{2\sigma_i^2} \right)
\end{align}
where $f(\bm{y}_\mathrm{obs}|\bm{\theta})$ is the likelihood of observing data $\bm{y}_\mathrm{obs}$ (Table \ref{tab:ecosystem_dataset}) from a model characterised by parameters $\bm{\theta}$, $N$ is the number of species which is $4$ for this system, $D$ is the number of observations per species which is $9$ for this system, $\sigma_i$ is a parameter representing measurement noise for species $i$, $y_{\mathrm{obs},i}(t_j)$ is the $j$th observation of species $i$, and $y_{\mathrm{sim},i(t_j)}$ is the data simulated by the model characterised by parameters $\bm{\theta}$ for species $i$ at the time point of the $j$th observation. 

We generated 100000 parameter sets for each of the target distributions using our SMC algorithm that blends exact and approximate approaches. Details on this sampling algorithm can be found in Supplementary Material Section \ref{SM: sampler}.

\subsubsection{Case study 3: Biochemical adaptation}
\label{SM:biochemical equations}

Section \ref{cellular adaptation case study}, \nameref{cellular adaptation case study} in the main text describes a biochemical reaction network used to model a system of interacting protein substrates ($A$ and $B$), their activated forms ($A^*$ and $B^*$), enzymes ($E_1$ and $E_4$), and intermediate protein-protein complexes ($C_1=[A E_1], C_2=[A^* B^*], C_3=[A^* B]$ and $C_4=[B^* E_4]$). In this paper, we use a general modelling framework known as the `complex complete framework', which explicitly accounts for the formation of intermediate complexes. The system of biochemical reactions depicted in Figure \ref{fig:Cellular main}A is described by the following ten ordinary differential equations \citep{jeynes_2023_protein_protein}:
\begin{align}
    \frac{\mathrm{d}A}{\mathrm{d}t} &= d_1C_1+k_2C_2-a_1AE_1, \\
    \frac{\mathrm{d}A^*}{\mathrm{d}t} &=k_1C_1+d_2C_2+d_3C_3+k_3C_3-a_3A^*B-a_2A^*B^*, \\
    \frac{\mathrm{d}B}{\mathrm{d}t} &= d_3C_3+k_4C_4 -a_3BA^*, \\
    \frac{\mathrm{d}B^*}{\mathrm{d}t} &= d_2C_2+k_2C_2+d_4C_4+k_3C_3-a_2A^*B^*-a_4B^*E_4, \\
    \frac{\mathrm{d}E_1}{\mathrm{d}t} &= (d_1+k_1)C_1 - a_1AE_1, \\
    \frac{\mathrm{d}E_4}{\mathrm{d}t} &= (d_4+k_4)C_4 - a_4B^*E_4, \\
    \frac{\mathrm{d}C_1}{\mathrm{d}t} &= a_1AE_1 - (d_1+k_1)C_1, \\
    \frac{\mathrm{d}C_2}{\mathrm{d}t} &= a_2A^*B^* - (d_2+k_2)C_2, \\
    \frac{\mathrm{d}C_3}{\mathrm{d}t} &= a_3A^*B^*-(d_3+k_3)C_3, \\
    \frac{\mathrm{d}C_4}{\mathrm{d}t} &= a_4B^*E_4 - (d_4+k_4)C_4. 
\end{align}
Here, $t$ is time, $a_i$ are the association rate constants, $d_i$ are the dissociation rate constants, and $k_i$ are the catalytic rate constants for $i=\{1,2,3,4\}$. For this study we fix parameters related to the total substrate $A_{\rm{TOT}}=A+A^*+C_1+C_2+C_3=10$ and $B_{\rm{TOT}}=B+B^*+C_2+C_3+C_4=10$. All 12 parameters in the model are highly uncertain and require calibration (see Table \ref{tab:biochemical parameters}). For more information on this model and its parameters, see \cite{jeynes_2023_protein_protein}. 

\begin{table}[H]
    \centering
    \begin{tabular}{p{0.3\linewidth}|p{0.3\linewidth}|p{0.2\linewidth}}
        \textbf{Parameter} & \textbf{Description} & \textbf{Prior} \\ 
        $a_i$ for $i=\{1,2,3,4\}$ & Association rates & $\mathrm{log} a_i \sim \mathcal{U}(-3, 4)$\\
        $d_i$ for $i=\{1,2,3,4\}$ & Dissociation rates & $\mathrm{log} d_i \sim \mathcal{U}(-3, 4)$\\
        $k_i$ for $i=\{1,2,3,4\}$ & Catalytic rates & $\mathrm{log} k_i \sim \mathcal{U}(-3, 4)$\\
    \end{tabular}
    \caption{The model parameters associated with the biochemical network model used in Section \ref{cellular adaptation case study}, \nameref{cellular adaptation case study}.}
    \label{tab:biochemical parameters}
\end{table}

To solve this system of ordinary differential equations, we simulate the system specified by the initial conditions in Table \ref{tab:biochemical initial conditions} for $t=0$ to $t=10^6$ and assume that the model has reached a steady state after this time. Then, the system is perturbed by increasing the concentration of $E_1$ by $1$, and the system is again solved for a period of $10^2$. 

\begin{table}[H]
    \centering
    \begin{tabular}{c|c}
        Initial condition & Value \\
        $A(0)$      & 10    \\
        $A^*(0)$    & 0     \\
        $B(0)$      & 10    \\
        $B^*(0)$    & 0     \\
        $E_1(0)$      & 1     \\
        $E_4(0)$    & 1     \\
        $C_1(0)$    & 0     \\
        $C_2(0)$    & 0     \\
        $C_3(0)$    & 0     \\
        $C_4(0)$    & 0     \\
    \end{tabular}
    \caption{Initial condition parameter values assumed in the biochemical network model used in Section \ref{cellular adaptation case study}, \nameref{cellular adaptation case study}.}
    \label{tab:biochemical initial conditions}
\end{table}

For the system to be considered adaptive, two conditions must be met after it is stimulated: \textit{sensitivity} and \textit{precision} \citep{ma_2009}. Sensitivity, $S$, measures the ability for a chemical concentration, $O$, to react to a change in stimulus, $I$,
\begin{align}
    S &= \frac{\left| O_{\rm{peak}}-O_1 \right| / O_1}{|I_2 - I_1|/I_1}, 
\end{align}
where $O_1$ and $I_1$ are the original chemical concentrations (output) and stimulus (input), respectively, $I_2$ is the updated stimulus, and $O_{\rm{peak}}$ is the concentration of $O$ which is furthest from $O_1$ after altering the stimulus (see Figure \ref{fig:Cellular main}B). A biochemical network that is sensitive to changes in input will have $S>1$ \citep{ma_2009}. 

The precision, $P$, is a measure of how well the system can return to its original value,
\begin{align}
    P &= \left( \frac{|O_2-O_1|/O_1}{|I_2-I_1|/I_1}\right)^{-1}, 
\end{align}
where $O_2$ is the final chemical concentration (output) after being stimulated. A biochemical network is considered precise if $P>10$ \citep{ma_2009}. Hence, we can define the discrepancy between adaptive biochemical networks and a model simulation as
\begin{align}
    \rho &=  \rm{max}(0,1-S) + \rm{max}(0,10-P) 
\end{align}
such that $\rho$ measures the discrepancy in sensitivity and precision for some stimulus.

We generated 10000 parameter sets that led to sensitive and precise biochemical systems using an SMC-ABC sampler \citep{drovandi_2011_ABC}. Of the 10000 parameter sets drawn naively from the prior distribution 0.18\% of these satisfied the precision and sensitivity requirements.  

In this case study, we are particularly interested in the Michaelis-Menten constants $K_3$ and $K_4$ which relate to the binding affinity of the complexes $C_3$ and $C_4$. These Michaelis constants are calculated using the association, dissociation and catalytic rates as 
\begin{equation}
    K_i=(d_i+k_i)/a_i,
\end{equation}
where $K_i$ is the $i$th Michaelis-Menten constant. Previous literature indicates that for this biochemical network to be capable of adaptation, both $K_3<1$ and $K_4<1$ \citep{ma_2009,jeynes_2023_protein_protein}.

\subsection{Additional results for each of the case studies}
\subsubsection{Case study 1: Logistic coral growth}
\begin{figure}[H]
    \centering
    \includegraphics[width=0.8\linewidth]{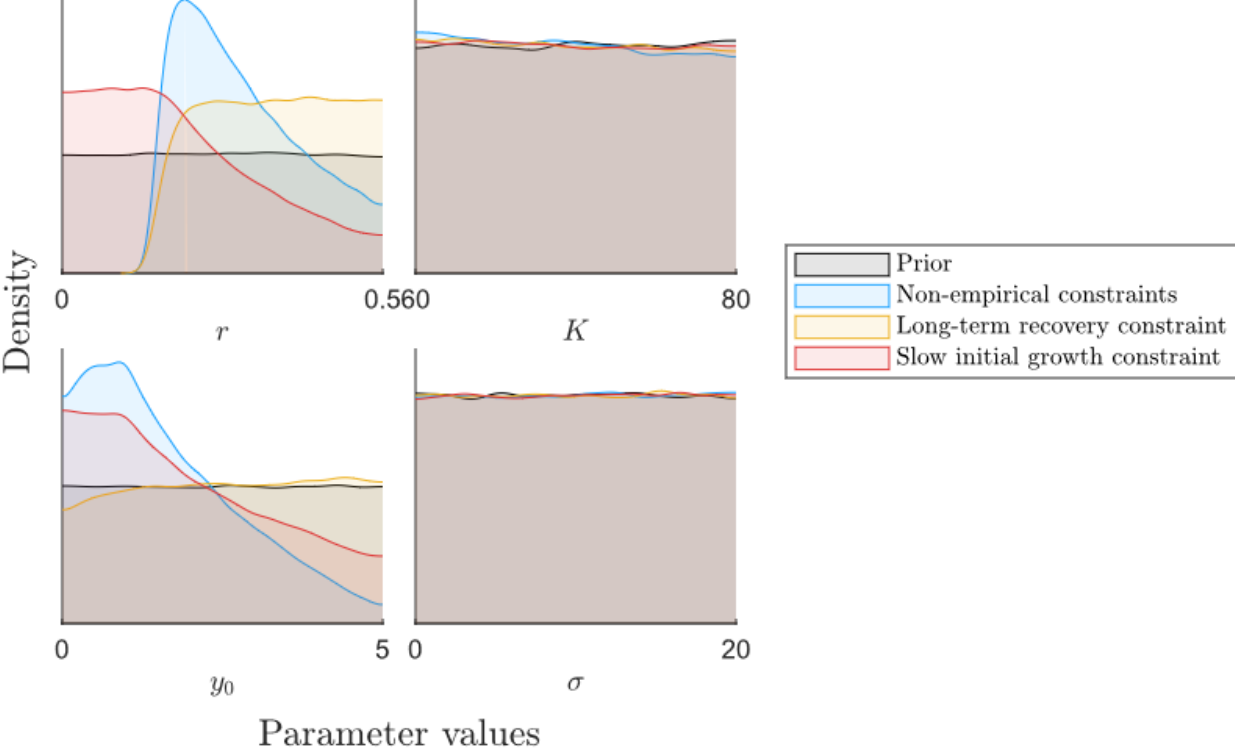}
    \caption{The marginal distributions of each model parameter for the logistic growth example from the prior, both non-empirical constraints individually and in combination. Notice that the distribution of the combination of two non-empirical constraints, is not necessarily the overlap between distributions. }
    \label{fig:Logistic_marginals}
\end{figure}

\begin{figure}[H]
    \centering
    \includegraphics[width=0.6\linewidth]{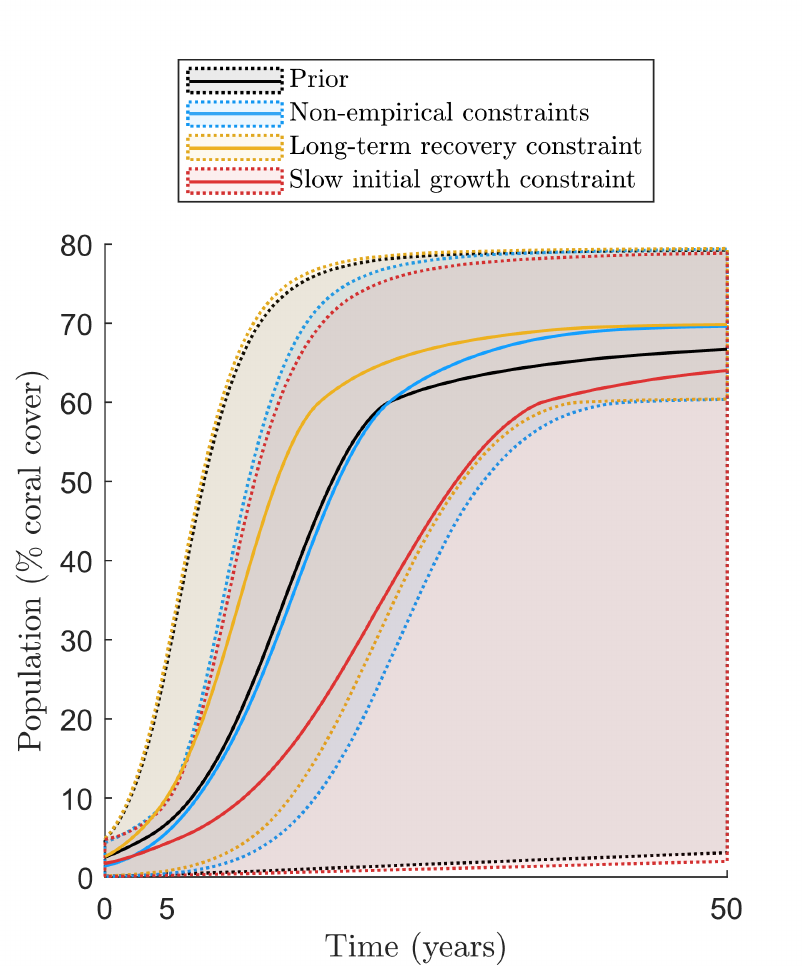}
    \caption{Model forecasts generated using ensembles of model parameters from the prior distribution (grey), the distribution with slow initial coral growth (red), the distribution with long-term coral recovery (yellow), and the distribution with both constrained (blue). Here we show the median prediction and the 95\% credible intervals. Notice that constraining the coral cover at 5 years mostly affects the earlier stages of predictions, and the constraint on 50-year coral cover mostly affects the late-stage predictions. Additionally, the overlap in predictions for each individual constraint matches the predictions from the joint distribution. }
    \label{fig:logistic_nonempirical_prediction_comparison}
\end{figure}

\begin{figure}[H]
    \centering
    \includegraphics[width=0.8\linewidth]{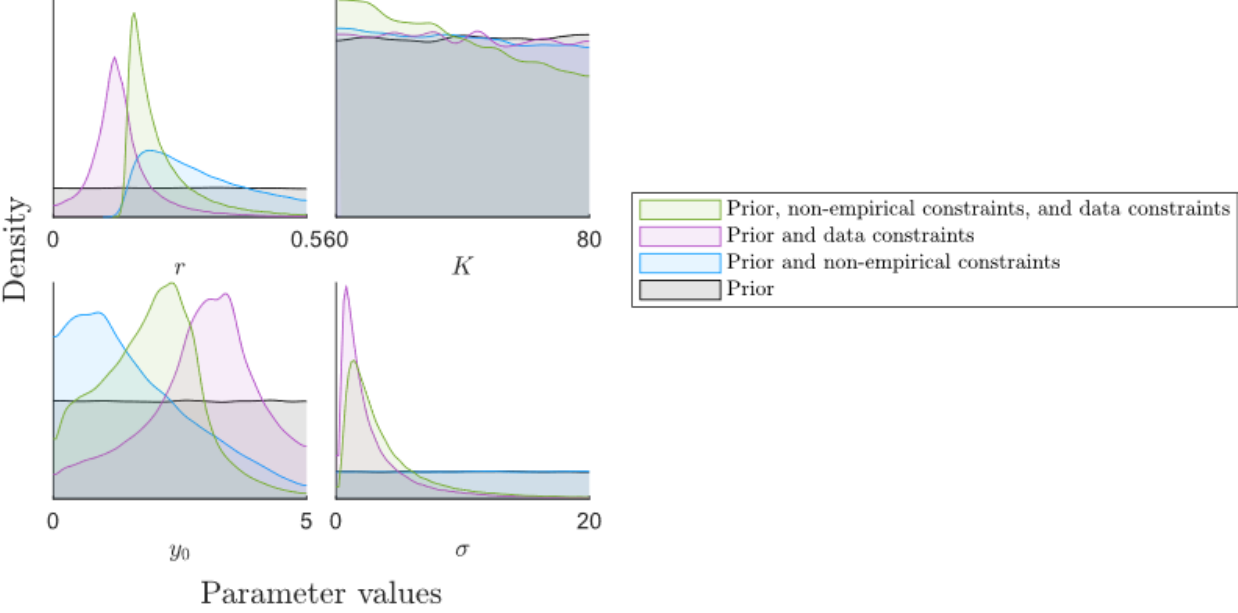}
    \caption{The marginal distributions of each model parameter for the logistic growth example from the prior, constraint posterior, data posterior, and constraint and data posterior distributions.}
    \label{fig:Logistic_marginals_data}
\end{figure}

\subsubsection{Case study 2: Ecosystem population modelling}

\begin{figure}[H]
    \centering
    \includegraphics[width=0.8\linewidth]{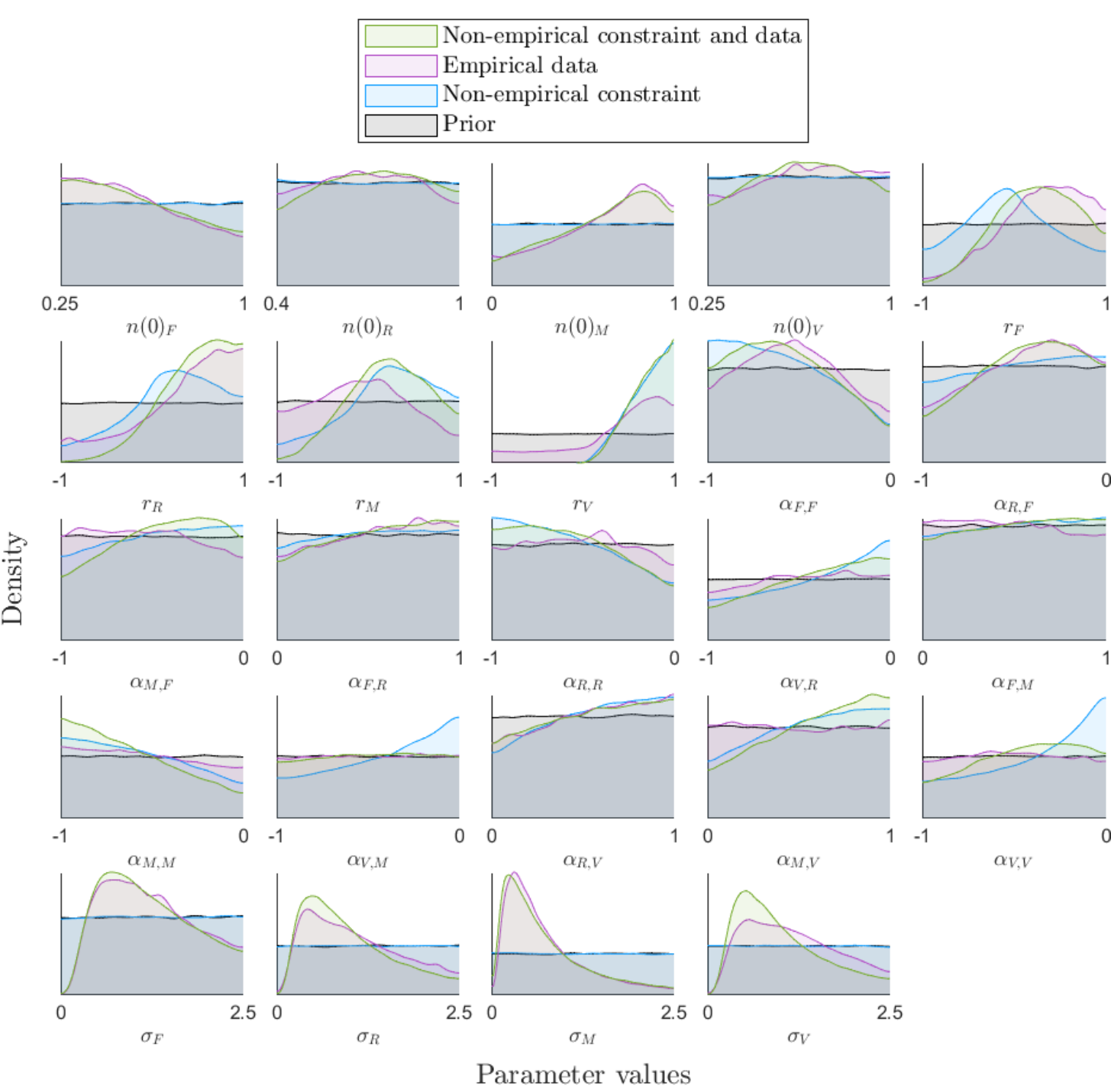}
    \caption{Marginal parameter distributions for each model parameter in the ecosystem population model. Four distributions of parameters are shown for the prior (grey), equilibrium-constrained posterior (blue), time-series data posterior (purple) and the combined equilibrium-constrained and time-series data posterior (green). Notice that the parameters most informed by the data are related to the Gaussian measurement noise $\sigma$. Whereas, the constraints on stability and coexistence yield far more informed parameter inferences.}
    \label{fig:ecol marginal}
\end{figure}

\begin{figure}[H]
    \centering
    \includegraphics[width=0.6\linewidth]{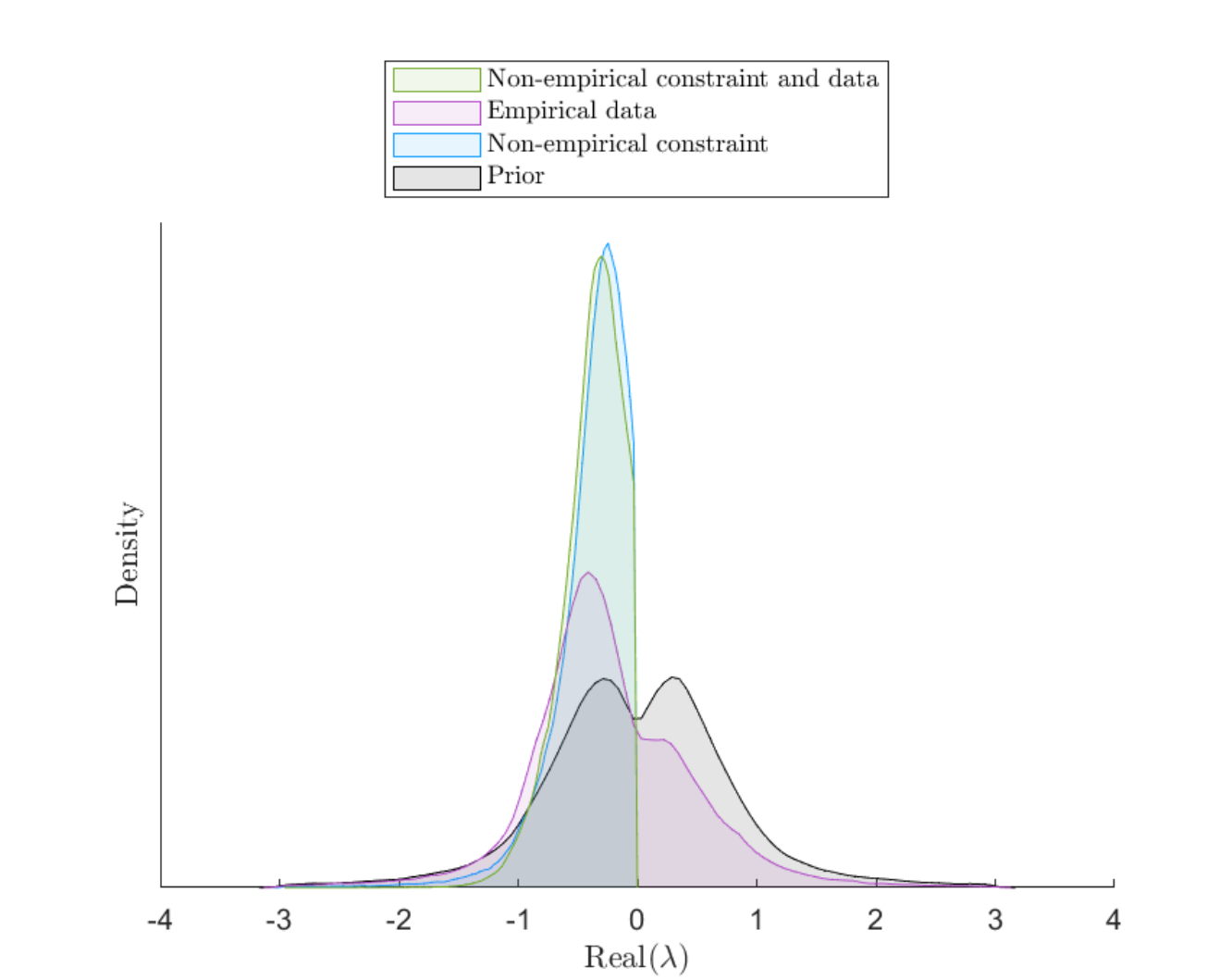}
    \caption{Distributions indicating the stability of the ecosystem population model. For each distribution of parameter sets, we show the distribution of the real parts of the eigenvalues of the Jacobian matrix evaluated at equilibrium, where negative values indicate stability.  Four distributions are shown for the prior (grey), equilibrium-constrained posterior (blue), time-series data posterior (purple) and the combined equilibrium-constrained and time-series data posterior (green). Notice that all distributions where stability is enforced are strictly negative (blue and green), and when informed by a dataset the system is more likely to be stable (purple) than when the data is not considered (grey). }
    \label{fig:ecol stability distribution}
\end{figure}

\subsubsection{Case study 3: Biochemical adaptation}

\begin{figure}[H]
    \centering
    \includegraphics[width=\linewidth]{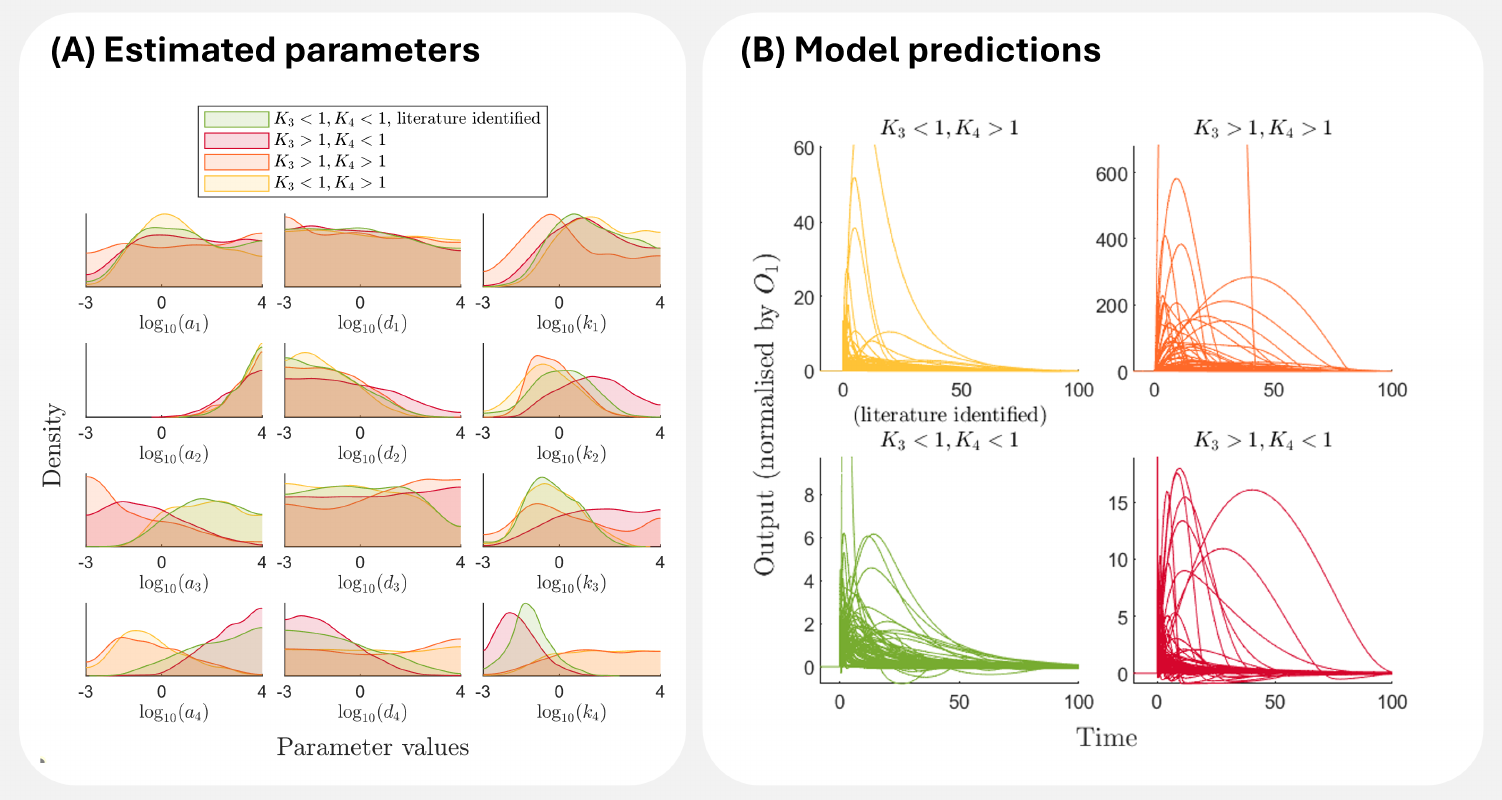}
    \caption{The parameter sets capable of adaptation, categorised according to their Michaelis-Menten constants $K_3$ and $K_4$ (see Figure \ref{fig:cellular K3vsK4} bottom left plot for these categories). \textbf{(A)} The estimated marginal parameter distributions for each of the Michaelis-Menten groups. \textbf{(B)} Model predictions for 100 randomly selected parameter sets from each Michaelis-Mentin group. Notice that both distributions and predictions for the literature-identified subgroup (green; $K_3<1,K_4<1$) are not distinct from the others, and that all plotted parameter sets meet the criteria of sensitive and precise adaptation. }
    \label{fig:cellular_K_group_comparison}
\end{figure}

\subsection{Sampling algorithm}
\label{SM: sampler}

Bayesian sampling algorithms aim to obtain a sample of parameter sets that is representative of the distribution being targeted \citep{martin_2020}. Sequential Monte Carlo (SMC) is one such sampling algorithm frequently used to obtain a sample from the posterior  \citep{delmoral_2006} or the approximate posterior distribution \citep{Sisson_2007_SMCABC}. SMC is a generally reliable sampling algorithm, and in particular, it is robust to the shape of the target distribution, minimises convergence issues, and is efficient when acceptance rates are low or the number of parameters is high \citep{Sisson_2007_SMCABC}.  

The general idea of an SMC algorithm is that an ensemble of weighted parameter sets is moved through a sequence of distributions, starting from the prior and ending at the target distribution, breaking the sampling problem into a series of simpler problems \citep{delmoral_2006}. For every sequence in the distribution, three steps are repeated -- reweighting, resampling, and moving -- until the final target distribution is obtained \citep{jeremiah_2012,chopin_2002,Beaumont_2019_ABC}. 
\begin{enumerate}
    \item \textbf{Reweighting:} Select a new intermediate target distribution and weight each parameter set using the probability of obtaining it from this distribution.
    \item \textbf{Resampling:} Duplicate and eliminate parameter sets depending on their weight. 
    \item \textbf{Moving:} Adjust the parameter values according to the intermediate target distribution via MCMC \citep{gamerman_2006}.
\end{enumerate}

When applied to a dataset, a likelihood function is used to weight the parameter sets, and a common choice for the sequence of distributions is likelihood annealing, whereby the intermediate target distributions are proportional to $\pi(\bm{\theta})f(\bm{y}_{\rm{obs}}|\bm{\theta})^\gamma$ where $\gamma$ is increased from $0$ (prior distribution) to $1$ (posterior). The effective sample size (ESS) is typically used when selecting $\gamma$ to ensure that not all of the weight is on very few parameter sets \citep{jasra_2011}. Similarly, for SMC-ABC, a discrepancy function is used to weight the parameter sets such that the intermediate target distributions are selected to iteratively reduce the discrepancy threshold $\epsilon$, where a common choice is to update the discrepancy threshold $\epsilon$ such that some percentage of parameter sets are below it \citep{drovandi_2011_ABC}. 

However, when targeting a distribution that includes both a likelihood and a discrepancy function (Equation \eqref{eq:exact and approximate posterior}), the weights on the parameter set need to be made up of both functions. Algorithm \ref{Alg:SMC} describes our SMC algorithm that simultaneously anneals both the likelihood and discrepancy information, where any changes from a standard SMC algorithm that only incorporates the likelihood are shown in blue. In this algorithm, the intermediate target distribution is determined by first selecting a new discrepancy threshold such that $a\%$ of parameter sets with the lowest discrepancy are retained, and all parameter sets with higher discrepancy are given zero weight. Then, the likelihood annealing temperature $\gamma$ is chosen using the updated weights, such that the influence of the likelihood is as high as possible, without dropping the effective sample size below some limit.  

In Algorithm \ref{Alg:SMC}, the tuning parameters used throughout this study were specified as $a=0.6, ESS_\mathrm{min}=0.3, c=0.99, n_\mathrm{MCMC}=10$. 

\begin{algorithm}[H]
    \small
    \BlankLine
    \textbf{INITIALISE} \\
    \BlankLine
    Import the dataset, $\bm{y}_\mathrm{obs}$ \\
    Specify the prior distribution, $\pi(\bm{\theta})$ \\
    Specify the likelihood function, $f(\bm{y}_\mathrm{obs}|\bm{\theta})$ \\
    {\color{blue}Specify the discrepancy function, $\rho(\bm{S}_\mathrm{obs},\bm{S}_\mathrm{sim}|\bm{\theta})$ \\}
    {\color{blue}Specify the target discrepancy threshold, $\epsilon$} \\
    Select the SMC tuning variables including: \\ 
    \qquad The number of particles to be sampled, $n$ \\     
    {\color{blue}\qquad The percentage of particles with the lowest discrepancy to be retained at each step, $a$ \\}
    \qquad The minimum acceptable ESS before resampling must occur, $\mathrm{ESS}_\mathrm{min}$, {\color{blue} where $\mathrm{ESS}_\mathrm{min} < a \times n$} \\
    \qquad The desired probability of particles moved during each MCMC step, $c$ \\
    \qquad The number of trial MCMC-ABC steps to gauge acceptance rate, $n_{\mathrm{MCMC}}$ \\ 
    \BlankLine
    Generate a sample of $n$ particles ($\{{\bm{\theta}}_i\}_{i=1}^{n}$) from the prior distribution, $\pi({\bm{\theta}})$ \\
    Set the particle weights ($\{W_i^t\}_{i=1}^n$) to be equal, such that $W_i^t=1/n$ \\
    Simulate the model for each $\{{\bm{\theta}}_i\}_{i=1}^{n}$, to calculate the log-likelihood $\log f(\bm{y}_\mathrm{obs}|\bm{\theta}_i)$ {\color{blue} and discrepancy $\rho(\bm{S}_\mathrm{obs},\bm{S}_\mathrm{sim}|\bm{\theta}_i)$} of each particle $i$\\
    Initialise the temperature for likelihood annealing, $\gamma_t=0$ \\
    {\color{blue}Calculate the maximum particle discrepancy, $\rho_{\mathrm{max}}=\mathrm{max}\{\rho(\bm{S}_\mathrm{obs},\bm{S}_\mathrm{sim}|\bm{\theta}_i)\}_{i=1}^n$} \\
    \BlankLine
    \BlankLine
    \While{{\color{blue}the discrepancy threshold is exceeded, $\rho_\mathrm{max}>\epsilon$, \textbf{OR}} the target distribution is not yet at the posterior, $\gamma_t<1$} {
    \BlankLine
    {\color{blue}Sort the particles by their discrepancy, $\rho(\bm{S}_\mathrm{obs},\bm{S}_\mathrm{sim}|{\theta})$ \\
    Select the discrepancy threshold $\epsilon_t = \rho(\bm{S}_\mathrm{obs},\bm{S}_\mathrm{sim}|{\theta}_{a \times n})$ \\
    Set the weights to 0 for particles with discrepancy above $\epsilon_t$, effectively removing those particles\\}
    Select the next temperature in the sequence, $\gamma_t$: \\
    \qquad \textbf{if} the posterior distribution ($\gamma_t=1$) has an acceptable sample size ($\geq \mathrm{ESS}_{min}$), \\
    \qquad \textbf{then} set $\gamma_t=1$. \\
    \qquad \textbf{else}, {\color{blue} using the updated weights $\{W_i^t\}_{i=1}^n$, }find the temperature at which \\ 
    \qquad \qquad $\mathrm{ESS}_t(\gamma_t)=\mathrm{ESS}_{min}$ where $\gamma_{t-1} < \gamma_t < 1$ using the bisection method\\
    \BlankLine
    \textbf{REWEIGHT} \\
    \BlankLine
    Reweight the sample to fit the current distribution, such that $w_i^t=W_i^{t-1}f(\bm{y}_\mathrm{obs}|\bm{\theta}_i^{t})^{\gamma_t-\gamma_{t-1}}$ \\
    Normalise the weights, $W_i^t=w_i^t/\sum_{k=1}^n w_k^t$ \\
    Calculate the current ESS, $\mathrm{ESS}_{t}=1/\sum_{i=1}^n(W_i^t)^2$
    \BlankLine
    \BlankLine
    \BlankLine
    \textbf{RESAMPLE} \\
    \BlankLine
    Resample particle values with probabilities given by their weights, $W_i^t$ \\
    Reset the particle weights $W$ such that $W_i^t=1/n$  \\
    Calculate the sample covariance matrix, $cov(\{\bm{\theta}_{i}\}_{i=1}^{n})$, to be used in the MCMC proposal distribution, $q_t$  \\
    \BlankLine
    \BlankLine
    \textbf{MOVE} \\
    \BlankLine
    \For{each of the $n_{\mathrm{MCMC}}$ trial MCMC steps}{
        Move the particles using MCMC (Algorithm \ref{Alg: MCMC})
        \BlankLine
    }
    Estimate the MCMC acceptance rate for iteration $t$, $a_t$ \\
    Determine the number of MCMC iterations to perform, $R_t = \lceil \log (c)/\log(1-a_t) \rceil$ and update $n_\mathrm{MCMC}=\lceil R_t/2 \rceil$ \\
    \For{each of the remaining MCMC steps, $R_t-n_{\mathrm{MCMC}}$}{
        Move the particles using MCMC (Algorithm \ref{Alg: MCMC})
        \BlankLine
    }
    \BlankLine
    \BlankLine
    {\color{blue}Update the maximum particle discrepancy, $\rho_\mathrm{max}$ \\}
    \BlankLine
    \BlankLine }
    \caption{SMC algorithm which incorporates both a likelihood and a discrepancy function. Differences from a standard SMC algorithm that only incorporates the likelihood are highlighted in {\color{blue} blue}.}
\label{Alg:SMC}
\end{algorithm}

\begin{algorithm}[H]
    \For{each particle $i$ in $\{\bm{\theta}_{i}\}_{i=1}^{n}$}
    {
        Propose a new set of parameter values  ${{\bm{\theta}_i}}^*$ using a multivariate normal proposal distribution, ${{\bm{\theta}_i}}^* \sim N({\bm{\theta}_i},cov(\{\bm{\theta}_{i}\}_{i=1}^{n}))$ \\
        Accept or reject the proposed particle value based on a Metropolis-Hastings acceptance probability, $\alpha = \min \left(1, \frac{\pi(\bm{\theta}^*_i)f(\bm{y}_\mathrm{obs}|\bm{\theta}^*_i)^{\gamma_t} {\color{blue}\mathbb{I}(\rho(\bm{S}_\mathrm{obs},\bm{S}_\mathrm{sim}|\bm{\theta}_i^*)<\epsilon_t)}}{\pi(\bm{\theta}_i)f(\bm{y}_\mathrm{obs}|\bm{\theta}_i)^{\gamma_t} {\color{blue}\mathbb{I}(\rho(\bm{S}_\mathrm{obs},\bm{S}_\mathrm{sim}|\bm{\theta}_i)<\epsilon_t)}} \right)$  
     } 
    \BlankLine
    Assess the portion of mutated particles \\
    \BlankLine
\caption{MCMC algorithm which incorporates both a likelihood and a discrepancy function used within the SMC algorithm (Algorithm \ref{Alg:SMC}). Differences from a standard MCMC algorithm that only incorporates the likelihood are highlighted in {\color{blue} blue}.}
\label{Alg: MCMC}
\end{algorithm}

\endgroup

\end{document}